\documentclass[10pt]{article}
\usepackage{fullpage}

\usepackage[hidelinks]{hyperref}
\usepackage[round, sort&compress]{natbib}
\usepackage{amsmath,amssymb}
\usepackage{amsthm}
\usepackage{mathtools}
\usepackage{verbatim}
\usepackage{subcaption}
\usepackage{booktabs}

\usepackage{float}
\usepackage[subnum]{cases}
\usepackage{graphicx}
\usepackage{algorithm}
\usepackage{algorithmicx}
\usepackage{algcompatible}
\usepackage{siunitx}
\usepackage{eucal}
\usepackage{caption}
\DeclareMathAlphabet{\mathcal}{OMS}{cmsy}{m}{n}   
\newlength{\thickarrayrulewidth}
\def\tsc#1{\csdef{#1}{\textsc{\lowercase{#1}}\xspace}}
\tsc{WGM}
\tsc{QE}
\tsc{EP}
\tsc{PMS}
\tsc{BEC}
\tsc{DE}

\begin{document}


\title{A Traffic Prediction-Based Individualized Driver Warning System to Reduce Red Light Violations}

\author{\normalsize Suiyi He, Maziar Zamanpour, Jianshe Guo, Michael W. Levin, Zongxuan Sun}



\maketitle

\begin{abstract}

\textcolor{black}{Running red lights remains a major cause of intersection crashes, injuries, and fatalities.}
Although numerous counter‑measures have been proposed, they continue to be a major problem in practice, partly because most existing systems deliver uniform guidance to every driver, prompting some motorists to ignore or misinterpret the alerts and leaving a persistent safety gap.
\textcolor{black}{
We present a novel method for providing accurate, individualized warnings in place of the broad, one-size-fits-all alerts used by most existing systems.}
Recognizing if a driver will run red lights is highly dependent on signal phase and timing, traffic conditions along the road, and individual driver behavior, the proposed warning system contains three parts.
First, a traffic prediction algorithm uses V2V and V2I data to predict traffic conditions along the approach to a signalized intersection.
Second, an optimization based algorithm that computes driver-specific warning that minimizes the risk of a red light violation, given the predicted traffic states and driver reaction model.
Third, the resulting advisory-quantifying the required deceleration-is presented on the in-vehicle display and updated continuously as the vehicle approaches the intersection.
Both numerical simulated driving scenarios and real-world road tests are used to demonstrate the proposed algorithm's performance under different conditions by comparing with previous work on red light running warning system.
The results show that the proposed system provides more effective and accurate warning signals to drivers.
In the simulation, the proposed algorithm cuts the ego vehicle's peak deceleration by up to 72.2\% relative to an unguided baseline, greatly reducing the risk of red-light violation.
\end{abstract}


\section{Introduction}\label{sec:rlrws-introduction}

Americans rely on road networks to transport people and goods through cities, but intersections, where vehicles travel in conflicting directions, present significant collision risks.
Traffic signals, using a simple green-yellow-red system, are a ubiquitous measure to prevent such collisions.
Nevertheless, red light running remains a frequent and often deadly problem.
According to the National Highway Traffic Safety Administration (NHTSA), between 2008 and 2019,  9,227 lives were lost in crashes involving red light running violations~\cite{nhtsa}, and this number continues to rise.
The record from the Insurance Institute for Highway Safety (IIHS) shows 1109 and 1149 deaths from red light running crashes in 2021 and 2022, respectively~\cite{iihs}.
Nearly half of these fatalities involve innocent pedestrians, cyclists, and passengers in vehicles hit by other red light runners.
Additionally, these violations caused approximately 127,000 and 107,000 injuries over the same two years.
The current scope of this problem means that additional systems are needed to reduce red light violations.

\subsection{Existing systems to reduce red light violations}

\textcolor{black}{
Several methods have been developed to warn drivers about red light violations or penalize them afterwards, but all existing systems have major shortcomings limiting their adoption and effectiveness at improving safety at signalized intersections.
{Red-light cameras} reduce the number of fatal red-light-running accidents by 21.3\%~\cite{hu2017effects}, and flashing green or yellow signals before the red light reduce red light violations~\cite{koll2004driver, mahalel1985safety, smith2001effects}. But both methods result in drivers braking aggressively at yellow lights; this harsh braking increases rear-end collisions by the following vehicles~\cite{huang2006effect, polders2015drivers, ahmed2015evaluation, claros2017safety}.}

\textcolor{black}{
{Countdown timers} indicating the remaining time had mixed results at reducing red light violations~\cite{long2013impact, long2011effects, ma2010investigating, chiou2010driver, sheykhfard2024exploring,yan2024exploring} due to aggressive drivers accelerating during yellow lights~\cite{fu2016effects}. Therefore, countdown timers sometimes increase the number of red light violations~\cite{long2011effects,biswas2017influence} and corresponding collisions because of the variety in driver responses. Similarly, pavement markings attempting to indicate stop/go decisions~\cite{elmitiny2010field,yan2007effect, yan2009impact} appeared to be effective at clarifying dilemma zones~\cite{zhang2014yellow} in simulation and driving experiments, but may encourage aggressive drivers to accelerate to attempt to enter the intersection. Lastly, instead of modifying driver behavior directly, some systems focus on adjusting signal timing strategies to enhance intersection safety~\cite{chen2021preventing, khalilabadi2024understanding}, extending yellow intervals~\cite{khalilabadi2024understanding} or all-red intervals~\cite{datta2000red,souleyrette2004effectiveness}.
However, drivers may adapt and become more aggressive about entering the intersection on yellow~\cite{liu2012empirical}, resulting in more red light violations in the long-term~\cite{souleyrette2004effectiveness}.}

\textcolor{black}{The shortcomings of these approaches can be traced to providing the same warning, penalty, or adjustment to all drivers (i.e. a ``broadcast''), encouraging a uniform and sometimes incorrect response. For example, red light cameras encourage harsh braking by drivers who should have entered the intersection on yellow, increasing rear-end collisions~\cite{wong2014lights}. Countdown timers encourage aggressive drivers to accelerate, when they should have stopped instead~\cite{fu2016effects}. Extending yellow and all-red intervals for everyone encourages drivers to attempt to enter the intersection on yellow when they should not~\cite{souleyrette2004effectiveness}.}


\subsection{Individualized red-light-running warning system}
\textcolor{black}{
The shortcomings of existing systems can be addressed by providing individualized guidance based on each driver's context (location, speed, signal timing, and traffic), ensuring appropriate stop or go decisions.} \textcolor{black}{
Rule-based warning systems have been widely adopted in prior studies by activating a warning signal to the driver. For example,~\cite{yan2015effect,johnson2019connected} assumed that the vehicle maintains a constant speed towards the intersection. 
While effective in simplified scenarios—where the ego vehicle is the only road user—this assumption neglects the influence of surrounding traffic, leading to inappropriate warnings in more complex settings.
Indeed, drivers often respond to such alerts with harsh braking~\cite{banerjee2020influence,zhang2021effect,zhang2022developing}, which can substantially increase the risk of rear-end collisions.
}

\textcolor{black}{
In contrast, we propose a novel framework that provides a range of warnings to individual drivers.
Our method tailors alerts not only to the ego vehicle’s position, speed, and remaining signal interval but also to the predicted evolution of traffic flow.
By aligning warnings with anticipated vehicle trajectories and future signal timings, this approach aims to reduce red-light violations and promote smoother, safer driver responses.} The range of warning signal shown to the driver varies from green ``normal driving'' to yellow ``standard braking'' to red ``full brake''.
A ``green'' signal indicates that the driver does not need to take specific action immediately, while ``yellow'' and ``red'' warning signals indicate that the driver should decelerate gradually or execute hard braking to avoid running a red light. A driver who initially ignores a moderate ``yellow'' warning would be shown increasingly harsh warnings to brake. The driver guidance is adjusted in real-time based on the driver's response, which affects their location and speed.

\begin{figure}
    \centering   
    \includegraphics[width=0.88\linewidth]{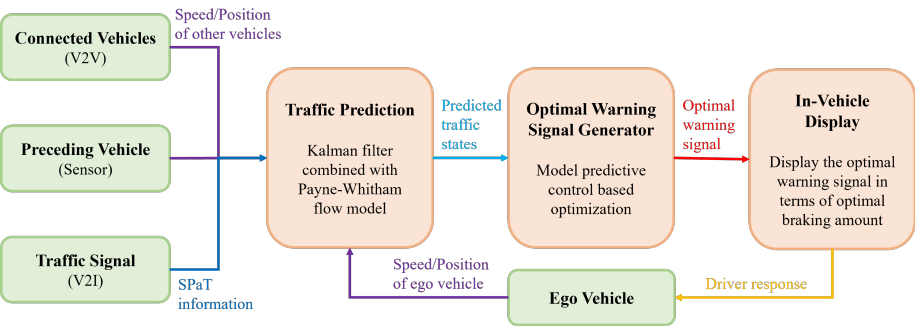}
    \caption{The structure of the proposed red light running warning system (RLRWS).}
    \label{fig:rlrws-structure}
\end{figure}

The overall design of our novel red light running warning system (RLRWS) is shown in Figure~\ref{fig:rlrws-structure}. 
\textcolor{black}{The driver receives visual indications of the optimal braking based on real-time traffic conditions and driving behavior.
This system is composed of several key components:
(1) A traffic prediction algorithm that predicts the future traffic conditions along the road section towards the signalized intersection using real-time information obtained from vehicle-to-vehicle (V2V) and vehicle-to-infrastructure (V2I) communications.
(2) An optimal warning signal generator that computes the optimal warning signal based on the predicted traffic conditions and driver model by optimizing the driver behaviour towards the signalized intersections.
By formulating the red light violation warning as an optimization problem, the system can mathematically account for various factors such as uncertainties in the driver model and different driving patterns, ensuring precise and adaptive alerts.
(3) An in-vehicle display that shows the generated warning signal to the driver in terms of optimal braking amount.
As the vehicle approaches the intersection, the proposed system continuously updates the warning signal based on the latest traffic condition and ego vehicle's status.
This ensures the system generates appropriate guidance by accounting for changes in surrounding driving conditions and variations in driver response to previous warning signals.}    

\textcolor{black}{
To obtain the inputs needed to calculate the optimal driver behavior, we assume that we have traffic signals equipped with V2I communications and a receiver in the ego vehicle. Furthermore, we sometimes assume that a subset of vehicles are equipped with V2V communications and are broadcasting their position and speed for the purposes of traffic prediction. 
}

\subsection{Contributions}

The contributions of this paper are as follows:
\begin{itemize}

    \item \textcolor{black}{We present a novel red light running warning system framework that proactively alerts drivers, helping them avoid red light violations.    
    By modeling the warning system into an optimization problem, the proposed framework has the capability to generate tailored warning signals to each individual driver.
    By using real-time traffic conditions obtained through vehicle-to-vehicle and vehicle-to-infrastructure communications, as well as the ego vehicle's speed and position, the system generates individualized warning signals, suggesting the optimal braking pattern for each driver to avoid running the red light while considering surrounding traffic evolution.}
    \item The system employs a traffic prediction algorithm to predict future traffic states based on the obtained real-time traffic information.
    A model predictive control based optimization framework is then used to compute the optimal warning signal.
    As the vehicle approaches the intersection, the system continuously updates its predictions and adjusts the warning signal, dynamically adapting to the changing traffic conditions and driver behavior. This approach offers a more refined and adaptive solution, addressing several key limitations found in existing red light violation countermeasures.
    
    \item We validate the effectiveness of our algorithm through both simulated traffic scenarios and real-world tests on public roads. 
    The results show that our approach is applicable to provide drivers with appropriate and timely warning signals across various driving scenarios, as shown in both numerical simulations and experiments.
\end{itemize}

The remainder of this paper is organized as follows. Section \ref{litreview} reviews related work on systems to reduce red light violations. In Section \ref{sec:rlrws-background}, we discuss traffic prediction to predict the future position of the target vehicle. Section \ref{sec:rlrws-algorithm} presents our novel algorithm. We demonstrate the algorithm in simulation in Section \ref{sec:sim-results} and on public roads in Section \ref{sec:rlrws-roadtest-results}. Finally, we conclude in Section \ref{sec:conclusion}.

\section{Related Work} \label{litreview}

\textcolor{black}{
The issue of red light violations has been addressed by many previous studies. Their approaches can be categorized into ``Conventional'' systems providing the same warning, penalty, or adjustment to all drivers due to the limitations of older technologies, and newer systems that attempt to provide in-vehicle warnings to individual drivers.
}

\subsection{Conventional red light enforcement systems}
\textcolor{black}{
Conventional red light enforcement systems—such as cameras, timers, and signal extensions—aim to improve safety but face limitations in driver behavior response, reliability, and real-world effectiveness.}

\textcolor{black}{
Red light cameras are reported to reduce the number of fatal red light running accidents by 21.3\%~\cite{hu2017effects} by issuing tickets to drivers who violate red lights. Although being effective, it can lead drivers to brake abruptly during the yellow interval to avoid fines, increasing the likelihood of rear-end collisions~\cite{huang2006effect, polders2015drivers, ahmed2015evaluation, claros2017safety}, and potentially increasing the total number of collisions~\cite{wong2014lights}. Additionally, unnecessary stops during the yellow interval can also reduce traffic flow efficiency, contributing to increased traffic congestion~\cite{hussain2020innovative}. Overall, despite their availability in practice, red light cameras are not used in many cities, and their legality varies by location~\cite{shannon2007speeding}. Flashing green or yellow signals have also shown an increased number of early stops~\cite{mahalel1985safety, smith2001effects,koll2004driver}. A related approach is using colored LED lights in the road to inform drivers on whether they should stop or enter the intersection on yellow~\cite{hussain2020improved, hussain2020innovative}.
However, such guidance ignores current vehicle speeds and future speeds limited by traffic conditions, as well as having visibility challenges in adverse weather conditions. Countdown timers on traffic signals, which display the remaining time for green and yellow intervals, are employed in many cities. 
\cite{johnson2019connected} proposed an in-vehicle countdown timer.
However, their impact on reducing red light violations is inconsistent in previous work~\cite{long2013impact, long2011effects, ma2010investigating, chiou2010driver, sheykhfard2024exploring,yan2024exploring, fu2016effects}.}

\textcolor{black}{
Signal controllers with dynamic yellow interval~\cite{khalilabadi2024understanding, arafat2023hardware} or all-red interval~\cite{zhang2011dynamic,park2016design,park2018field,simpson2023dynamic} allow red light runners to fully cross the intersection before vehicles from other directions proceed, and also prevent vehicles from entering the dilemma zone~\cite{bonneson2002intelligent, tarko2006probabilistic}.
To ensure a reliable signal output, algorithms must accurately predict drivers' stop and go decisions, using learning-based~\cite{li2014red,jahangiri2015adopting,huang2015redeye} and probability-based~\cite{zhang2011dynamic,tan2018real} approaches with a large amount of data set requirement for every implemented intersection, and rule-based~\cite{park2016design, park2018field,simpson2023dynamic} approaches. Although the rule-based approach in \cite{park2018field} reported a 100\% detection rate, it exhibits an up to 30\% false alarm rate, potentially reducing traffic flow efficiency due to unnecessary all-red extensions.}

\textcolor{black}{
Overall, conventional red light enforcement systems rely on generalized warnings or control strategies that fail to account for individual driver behavior and real-time traffic dynamics. This can lead to two extremes: drivers ignoring the message or reacting abruptly, causing rear-end collisions or reduced traffic efficiency. These drawbacks highlight the need for adaptive, individualized messaging that balances safety and flow by aligning guidance with each driver's specific context.}

\subsection{Individual driver guidance}
To warn drivers about crossing the stop bar during a red light, studies such as \cite{yan2015effect,johnson2019connected, banerjee2020influence,gelbal2020hardware, hadi2021evaluating, zhang2021effect,zhang2022developing,tajalli2022testing, dokur2022vehicle} propose in-vehicle warning systems that alert drivers, helping to prevent red light violations and avoid conflicts with other road users. These systems usually use connected vehicle-to-infrastructure data to obtain future signal timings.
Results from driving simulators, field tests and road tests demonstrate these systems' effectiveness in stopping red light runners. 
However, these methods typically rely on rule-based approaches, where the decision to issue a warning is based solely on the ego vehicle’s real-time speed, position, and SPaT data, assuming constant speed or deceleration. 
This approach neglects the impact of surrounding traffic dynamics and the behavior of the ego vehicle’s driver on red light violations. \textcolor{black}{
However,  the in-vehicle warning system with a single stage visual warning (such as warning lights) can lead to aggressive braking when the warning first appears~\cite{banerjee2020influence}.
This issue may still exist in two-stage warning systems~\cite{zhang2021effect,zhang2022developing} as the sudden appearance of the warning message can cause drivers to brake sharply at the initial stage.}

To reduce the impact of red light violations on other vehicles, \cite{xiang2016effect,zhang2015discrimination} propose in-vehicle warning systems that alert drivers near signalized intersections about other vehicles running a red light.
While these systems may reduce the risk of angle collisions caused by red light runners, they do not prevent the violations themselves.

\subsection{Optimal vehicle control}
The use of autonomous driving technology to improve intersection safety is studied in~\cite{al2023self,lu2023game,zhao2023multi}.
However, \textcolor{black}{as fully autonomous driving systems are still in the development stage, improving intersection safety at the current stage requires methods to reduce red light violations by human drivers. Among preventive available approaches,} 
For example, \cite{naik2020dilemma, naik2023evaluation} present a warning system that avoids falling into the dilemma zone considering the vehicle position and signal timing. 
\textcolor{black}{
Similarly, \cite{mahbub2022safety} utilizes vehicle connectivity in order to predict dilemma zones and warn the driver based on a model predictive control. Moreover, some studies have proposed trajectory smoothing methods to optimize vehicle acceleration near signalized intersections, aiming to reduce fuel consumption and improve safety~\cite{yao2018trajectory, wu2023joint}. While similar to our approach in providing real-time individualized speed control, these methods focus on smooth deceleration for automated vehicles rather than preventing red light violations.}

\textcolor{black}{
In summary of this section, designing an effective red light running warning system requires a framework capable of proactively alerting drivers while accounting for traffic dynamics and individual driver behaviour.
The warning signal must adapt gradually to changing traffic conditions, avoiding abrupt displays that may trigger sudden, harsh braking, otherwise resulting in either being ignored by the drivers due to being so generic, or showing a harsh brake with rear-end collision impact.}
Moreover, the warning should strike a balance, neither causing overly conservative nor excessively aggressive driving, ensuring that drivers maintain normal behavior while preserving intersection efficiency and safety.
To address these needs, we propose a novel framework that optimally delivers warning signals to drivers while considering driver's behaviour and the evolution of traffic states.
Using vehicle connectivity, traffic prediction, and optimization, our framework overcomes the limitations of existing systems.

\section{Algorithm Design}\label{sec:rlrws-algorithm}
We now present our RLRWS algorithm.
Our system consists of a MPC-based optimization problem, which is used to compute the optimal warning signal shown to the driver.

\subsection{RLRWS Framework}\label{sec:rlrws-framework}

The primary objective of the RLRWS is to prevent the driver from running red lights by warning the driver about required braking when appropriate.
The structural framework of our innovative system is illustrated in Figure~\ref{fig:rlrws-structure}.
As analyzed in previous sections, a driver's future actions while approaching a signalized intersection are influenced by future traffic conditions, which include both traffic flow dynamics and SPaT data.
Crucially, the warning system must accurately determine the ego vehicle's position in relative to the intersection as the traffic signal turns red.
Furthermore, in car-following scenarios, the behaviour of the ego vehicle must adapt to that of the immediate preceding vehicle, which affects the decision of how to respond to a yellow traffic light.
Predicting the longitudinal trajectory of this preceding vehicle is therefore essential for determining the appropriate warning.
To address these challenges, our proposed warning system employs the prediction framework shown in Section~\ref{sec:rlrws-background}, predicting the longitudinal movement of both the ego vehicle and its immediate preceding one as they approach the signalized intersection.

\textcolor{black}{
The purpose of the algorithm is to give appropriate  guidance on the braking required to stop at a red light, or suggest normal driving to enter the intersection on yellow.}
Therefore, the algorithm should compute the optimal driver behavior as the vehicle approaches a signalized intersection.
After predicting the future traffic states, the challenge of preventing red light violations is modeled as an optimization problem.
As previously discussed, the driver's future actions are constrained by the dynamics of the traffic flow.
Therefore, we formulate this optimization problem using a model predictive control (MPC) approach, which computes the optimal inputs for the system while accounting for various constraints along the prediction horizon.
MPC optimizes the system's input based on its predicted future states, making it particularly suitable for developing the RLRWS.
This approach allows the algorithm to consider the driver's anticipated behaviour in response to to future traffic conditions.
Since the MPC continuously updates its calculations, it reflects any changes, such as the driver's previous responses or evolving traffic conditions, in subsequent optimizations.
These continuous updates ensure that the warning signal are seamlessly adjusted, avoiding the abrupt triggers that could cause aggressive braking, as seen in the previous work.
In this way, an optimal, continuous warning signal can be generated.

Finally, the warning signal is displayed to the driver via an in-vehicle screen. 
As the vehicle approaches the intersection, the prediction algorithm updates the predicted traffic states every \SI{0.2}{\second}, while the optimization problem updates the optimal warning signal every \SI{1}{\second}, which is then reflected on the display.
These continuous updates ensures that the system delivers an accurate and timely warning, even in the event of unforeseen traffic conditions or unexpected driver behaviours as the vehicle approaches the intersection.

\subsection{MPC-based Optimization Problem}\label{sec:rlrws-mpc}
We formulate the problem of finding the optimal deceleration as a MPC-based optimization problem. By solving this repeatedly in real-time, we will adapt to the ego vehicle's location and speeds, and any changes caused by the driver.
This formulation is based on traffic prediction results discussed in Section \ref{sec:rlrws-background}.
The mathematical formulation of the optimization problem is as follows: 
\begin{subequations}
\label{eq:rlrws-warning-mpc}
\begin{align}
    u^{*}(\cdot)~{=}~\text{argmin} \int^{t_f}_{t_0}&q(x(t),v(t),a(t))dt\label{eq:rlrws-mpc-cost}\\
    \text{s.t.} \quad 
    \Dot{x}(t) = v(t), &\label{eq:rlrws-dynamics-pos} \\
    \Dot{v}(t) = a(t), &\label{eq:rlrws-dynamics-spd}\\
    a(t) = f(u(&t), x(t), v(t)), \label{eq:rlrws-mpc-driver-model}\\
     u(t) \in \mathcal{U},~~~& \label{eq:rlrws-mpc-constraint-warning}\\
    v_{\text{min}}\leq v(t) &\leq v_{\text{max}},~ a_{\text{min}}\leq a(t) \leq a_{\text{max}} ,\label{mpc:rlrws-constraint}\\
    x(t) \leq x_{\text{tl}}-&v(t)\tau_\text{tl},\text{ if the traffic light is red at $t$,}\label{mpc:rlrws-tl-constraint}\\
    v(t_f) = 0 \text{ a}&\text{nd }x(t_f)\geq x_{\text{tl}} - d_\text{tl}, \text{ if the traffic light is red at }t_f
    \text{ and }\hat{x}(t_f)\geq x_{\text{tl}}-d_\text{tl}, \label{mpc:rlrws-terminal-constraint}\\
    x(t) \geq \hat{x}_\text{lead}&(t) +\beta \sigma[\hat{x}_\text{lead}(t)] - d_\text{max}, \label{mpc:rlrws-max-spacing}\\
    x(t) \leq \hat{x}_\text{lead}&(t) - \beta \sigma[\hat{x}_\text{lead}(t)] - (d_\text{min} + h_\text{min}v(t)),\label{mpc:rlrws-min-spacing}\\
    x(t_0) = x(t_0&),~ v(t_0) = v(t_0), \label{eq:rlrws-mpc-initial-condition}
\end{align}
\end{subequations}
\textcolor{black}{where \eqref{eq:rlrws-mpc-cost} is the optimization problem's objective function; 
$x(t)$, $v(t)$ and $a(t)$ are the ego vehicle's longitudinal position, longitudinal speed and longitudinal acceleration, respectively;
$t_0$ and $t_f$ represent the beginning and end time step of the optimization problem.
\textcolor{black}{Details of this formulation will be discussed in the next subsection.}
\eqref{eq:rlrws-dynamics-pos} and \eqref{eq:rlrws-dynamics-spd} describes the ego vehicle's longitudinal dynamics in term of its longitudinal location and speed.
\textcolor{black}{
The function $f(\cdot)$ in \eqref{eq:rlrws-mpc-driver-model} represents the driver model.
Beyond capturing the driver's response to the warning signal $u(t)$ in the context of traffic conditions-such as the ego vehicle's speed $v(t)$ and position $x(t)$-this model can also incorporate individual-specific reaction characteristics, including reaction time, slow-down pattern.
By integrating this driver model into the proposed optimization framework, the generated warning signals are tailored to each driver, offering a more customized alternative to existing broadcast systems like countdown timers.
}
\eqref{eq:rlrws-mpc-constraint-warning} represents the constraint on the warning signal's value.
\eqref{mpc:rlrws-constraint} indicates the physical constraint on vehicle's speed and acceleration.
Constraints \eqref{mpc:rlrws-tl-constraint} and \eqref{mpc:rlrws-terminal-constraint} are used to stop the ego vehicle timely during the red light, will be talked in details in Section 4.2.3.
Inequalities \eqref{mpc:rlrws-max-spacing} and \eqref{mpc:rlrws-min-spacing} serve as the maximum and minimum spacing constraints between the ego vehicle and this immediate preceding one.
$\eqref{eq:rlrws-mpc-initial-condition}$ specifies the initial condition of the optimization problem.}

\textcolor{black}{In constraints \eqref{mpc:rlrws-tl-constraint} and \eqref{mpc:rlrws-terminal-constraint}, $x_{\text{tl}}$ represents the longitudinal position of the traffic light;
$\tau_\text{tl}$ is a newly introduced variable and is called the desired time headway for a red light;
$d_\text{tl}$ is a small buffer distance;
The variable $\hat{x}(t_f)$ denotes the ego vehicle's predicted longitudinal position at the terminal time step $t_f$ of the optimization problem.}

\textcolor{black}{In constraints \eqref{mpc:rlrws-max-spacing} and \eqref{mpc:rlrws-min-spacing}, $\hat{x}_\text{lead}(t)$ represents the predicted longitudinal position of the immediate preceding vehicle, obtained from the traffic prediction algorithm;
$d_{max}$ is the maximum following distance;
$d_{min}$ represents the minimum spacing;
$h_{min}$ denotes the time headway and is set to \SI{1.5}{\second} in this work;
$\sigma$ and $\beta$ represent the standard deviation and confidence level of the prediction, respectively.
The use of these constraints adapts to different conditions.}

\subsubsection{Objective Function}\label{sec:rlrws-mpc-objective-function}

Our objective function is $q(x(t), v(t), a(t)) = w_1 a^2(t) + w_2 \Dot{a}^2(t) + w_3 (v(t)-v_{\text{ref}})^2$, where $w_1$, $w_2$ and $w_3$ are weighting factors
and $v_{\text{ref}}$ is a reference longitudinal speed when the vehicle approaches a signalized intersection. The purpose of the objective is to minimize the vehicle's acceleration and jerk while tracking the reference speed along the prediction horizon,resulting in smoother longitudinal maneuvers.
\textcolor{black}{The reference longitudinal speed $v_{\text{ref}}$ is the desired speed that the driver should follow in the ideal condition.
When the traffic light is green, the reference speed is the free flow speed of the road.
When the traffic light is red, the reference speed gradually drops to zero before the stop bar.
To simplify the computational process of the solver, this reference speed is modeled using a sigmoid function.}

\subsubsection{Driver Model}\label{sec:rlrws-driver-model}

Given a warning signal $u(t)$, the driver responds by adjusting the vehicle's longitudinal speed $v(t)$.
However, the driver's reaction to the same warning signal varies depending on the vehicle's speed, position, and surrounding traffic conditions.
For instance, given the same warning signal, when the vehicle is farther from a signalized intersection, the driver tends to apply less force to the braking pedal compared to when the vehicle is closer to the red light.
Therefore, the actual relationship between the warning signal and the resulting vehicle acceleration depends on several factors.
In this work, the function $f(\cdot)$ is used to describe this relationship. \textcolor{black}{Any reasonable function that can estimate braking behavior of the driver can be integrated into our proposed framework.
For preliminary testing of our proposed warning system framework, a linear driver model $a(t) = -u(t)/20$ is used as this function $f(\cdot)$. However, this function can be substituted with a more detailed driver model that describes human responses better, which requires an extensive data recording and calibration.}

\textcolor{black}{Although real drivers exhibit complex, nonlinear responses to warning signals and traffic conditions—and may deviate from our simplified linear driver model—this approximation remains effective for computing appropriate alerts within our framework.
Since the MPC optimization problem continuously updates the warning signal every 1 second, using the most recent vehicle states as its initial condition, each updated warning signal inherently accounts for any non-adherence to the previous command.
For instance, if the warning signal desires a moderate 30\% braking but the driver does not actually brake, the system will generate a larger warning signal as the vehicle nears the intersection.}

Preventing drivers from red light violations typically requires them to decelerate before reaching the intersection.
In this work, we define the warning signal as a continuous scale ranging from 0 and 100 to represent the optimal braking intensity.
However, to adhere to car-following constraints and ensure smooth traffic flow, the optimal driver action may sometimes involve acceleration rather than deceleration, resulting in a positive $a(t)$ in \eqref{eq:rlrws-warning-mpc}.
Thus, to allow for the ego vehicle's acceleration, the value of $u(t)$ can also be negative, with a lower limit set at $-20$.
Then, the magnitude of the warning signal is visually represented in three colors--green, yellow and red--each varying in size according to the signal's intensity.

\subsubsection{Red Light Constraints}\label{sec:rlrws-traffic-constraints}

Each time the prediction algorithm updates the ego vehicle's predicted longitudinal trajectory along the prediction horizon, the warning system first determines whether the ego vehicle will cross the stop bar as the traffic signal turns red.
If the ego vehicle is predicted to pass the intersection at that moment, no warning is necessary.
Both of constraints \eqref{mpc:rlrws-tl-constraint} and \eqref{mpc:rlrws-terminal-constraint} are not needed.

When the ego vehicle is predicted to be unable to cross the stop bar at the traffic light turns red, the constrain \eqref{mpc:rlrws-tl-constraint} is used to prevent red light violations.
\textcolor{black}{The term $\tau_\text{tl}$ is introduced to the proposed warning framework and is called desired time headway for a red light.
This ensures that there is sufficient time for the ego vehicle to come to a smooth stop before the stop bar.}

As the ego vehicle gets close to the intersection, the system further evaluates whether the ego vehicle is predicted to reach the stop bar under a red light by the end of the prediction horizon $t_f$.
Specifically, in this work, we do this by determining whether the ego vehicle is predicted to be within $d_\text{tl}$ \SI{20}{\meter} of the stop bar at the terminal horizon $t_f$.
If this condition is met, the constraint \eqref{mpc:rlrws-terminal-constraint} will be incorporated into the optimization problem to ensure that the vehicle stops close to the stop bar at time $t_f$.
To avoid infeasibility, positive slack variables $\gamma_v$ and $\gamma_x$ are introduced to the optimization problem. 
Consequently, the constraint \eqref{mpc:rlrws-terminal-constraint} is rewritten as: $v(t_f) = \gamma_v \text{ and }x(t_f) + \gamma_x\geq x_{\text{tl}} - d_\text{tl}$.
The penalty cost terms $w_v \gamma_v^2 + w_x \gamma_x^2$, accompanied with two large positive weighting factors $w_v$ and $w_x$, are added to the objective function \eqref{eq:rlrws-mpc-cost}.
This strategy allows the optimization problem to slightly violate these constraints when they cannot be satisfied, thereby guaranteeing the feasibility of the algorithm.

As the ego vehicle approaches the stop bar and the traffic light remains red, a shorter prediction horizon is required by the MPC optimization, promoting a stricter formulation of the the terminal constraint \eqref{mpc:rlrws-terminal-constraint} regarding distance.
This adjustment is necessary because as the vehicle gets closer to the intersection, less time is needed to reach the stop bar, and the vehicle is expected to stop closer to it.
In this problem, the initial prediction horizon is set to \SI{10}{\second} and $d_\text{tl}$ is set to \SI{20}{\meter}.
As the ego vehicle approaches the stop bar, these values are updated as follows:
when the ego vehicle is \SI{60}{\meter} away from the stop bar, these values are updated to \SI{10}{\second} and \SI{15}{\meter};
when the ego vehicle is \SI{40}{\meter} away from the stop bar, these values are updated to \SI{8}{\second} and \SI{10}{\meter};
when the ego vehicle is \SI{20}{\meter} away from the stop bar, these values are updated to \SI{6}{\second} and \SI{5}{\meter}.
This process speeds up the computational process, ensuring timely and accurate adjustments as the vehicle approaches the intersection.

Meanwhile, different combinations of the car-following constraints \eqref{mpc:rlrws-max-spacing} and \eqref{mpc:rlrws-min-spacing} will be used in the MPC formulation, depending on the predicted trajectories of both the ego vehicle and its immediate preceding vehicle.
For clarity, we divide these combinations into two distinct categories:
\begin{itemize}
    \item \textbf{ Without Preceding Vehicle}: 
    In the absence of an immediate preceding vehicle ahead of the ego vehicle, the constraints related to traffic lights \eqref{mpc:rlrws-tl-constraint}--\eqref{mpc:rlrws-terminal-constraint} are the only constraints on the ego vehicle's longitudinal position.
    As previously mentioned, when the traffic light is red, the constraint \eqref{mpc:rlrws-tl-constraint} becomes activate.
    If the prediction algorithm indicates that the ego vehicle can reach the stop bar under a red light by the end of the prediction horizon $t_f$, the constraint \eqref{mpc:rlrws-terminal-constraint} is deployed to ensure the vehicle stops close to the stop bar.
    
    \item \textbf{With Preceding Vehicle}:
    When an immediate preceding vehicle is present ahead of the ego vehicle, and its speed and position data are accessible through onboard sensors or V2I communication, the constraints \eqref{mpc:rlrws-max-spacing}--\eqref{mpc:rlrws-min-spacing} on the spacing between the ego vehicle and this preceding one may be necessary.
    The constraint \eqref{mpc:rlrws-max-spacing} limits the maximum spacing, thereby ensuring mobility of the traffic flow.
    The constraint \eqref{mpc:rlrws-min-spacing} limits the minimum following distance, thereby guaranteeing the ego vehicle's safety.

    To predict the immediate preceding vehicle's longitudinal movement, the traffic prediction algorithm from \cite{he2023real} is used.
    Based on the relative position of the ego vehicle, its preceding one, and the stop bar at the moment the signal turns red, we consider two different scenarios:
    (1) If both the ego vehicle and its immediate preceding vehicle are predicted to not cross the stop bar when the red light appears, both spacing constraints \eqref{mpc:rlrws-max-spacing}--\eqref{mpc:rlrws-min-spacing} must be satisfied along the entire prediction horizon to ensure the mobility and safety of the traffic flow.
    In this case, the constraint \eqref{mpc:rlrws-terminal-constraint} becomes unnecessary, as the presence of the immediate preceding vehicle inherently constraints the stop position of the ego vehicle through the constrains on maximum and minimum spacing.
    (2) If only the immediate preceding vehicle is expected to pass through the stop bar before the traffic light turns red, only the minimum spacing constraint \eqref{mpc:rlrws-min-spacing} is included in the optimization to guarantee the safety.
    Similar to the scenario with no preceding vehicle, the constraint \eqref{mpc:rlrws-terminal-constraint} will be included in the formulation only if the ego vehicle is predicted to reach the stop bar by the terminal horizon $t_f$ while the traffic light is red.
\end{itemize}

\subsection{Warning Signal Visualization}\label{sec:rlrws-visualization}
\begin{figure}
    \centering
    \captionsetup{justification=centering}
    \begin{subfigure}[t]{0.32\textwidth}
    \centering
        \includegraphics[width=0.7\linewidth]{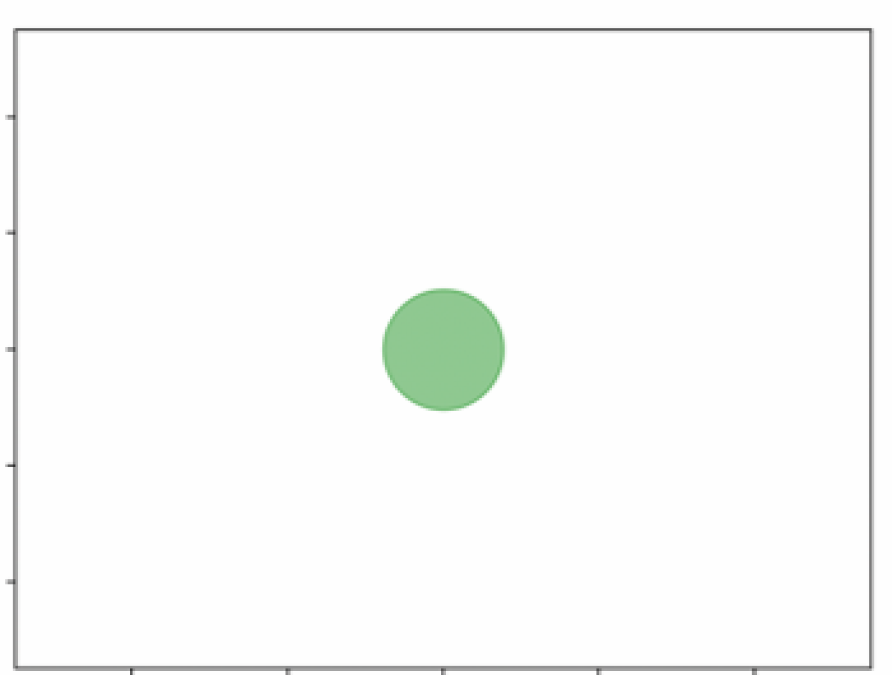}
        \label{fig:rlrws-warning-green}
        \caption{Example of green warning signal.}
    \end{subfigure}
    \begin{subfigure}[t]{0.32\textwidth}
    \centering
        \includegraphics[width=0.7\linewidth]{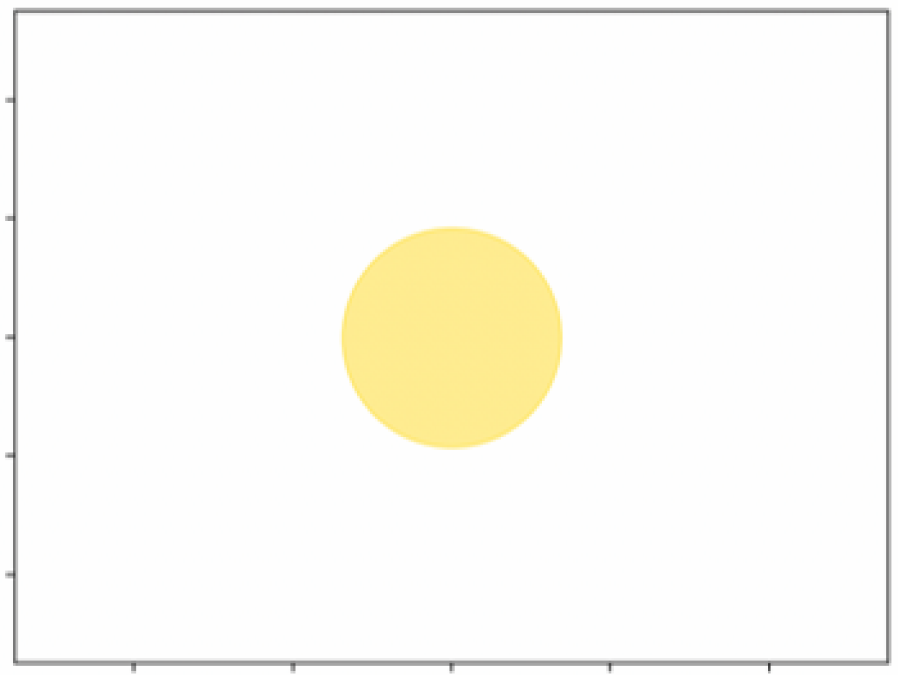}
        \label{fig:rlrws-warning-yellow}
        \caption{Example of yellow warning signal.}
    \end{subfigure}
    \begin{subfigure}[t]{0.32\textwidth}
    \centering
        \includegraphics[width=0.7\linewidth]{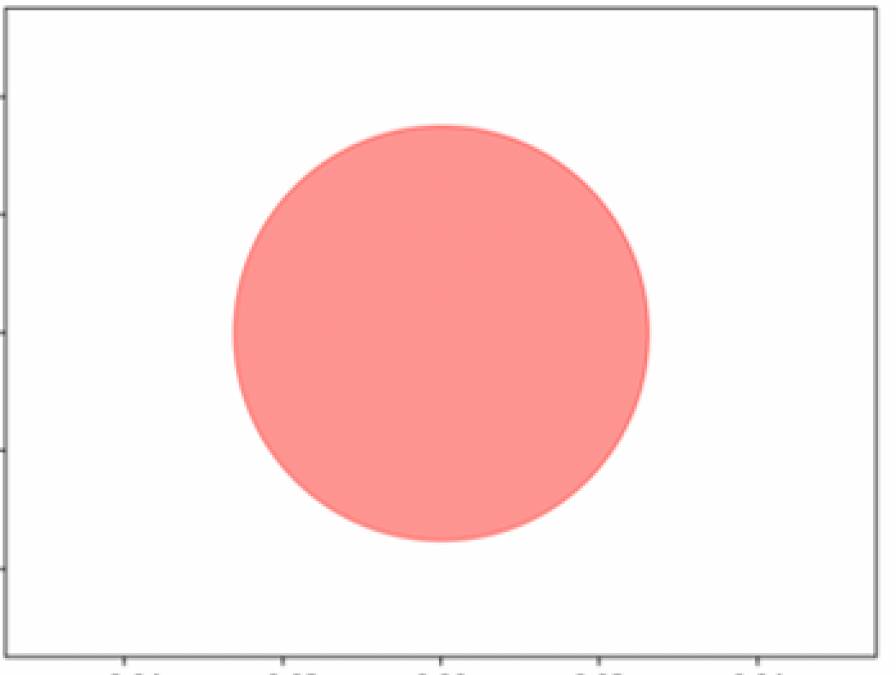}
        \label{fig:rlrws-warning-red}
        \caption{Example of red warning signal.}
    \end{subfigure}
    \caption{Warning signal visualization.}
    \label{fig:rlrws-warning-color}
\end{figure}

As mentioned in the previous subsection, the computed optimal warning signal is represented by three distinct colors—-green, yellow, and red—-each with varying sizes. 
This work employs a visual warning system featuring colored circles (Figure~\ref{fig:rlrws-warning-color}), where the diameter signifies the required braking intensity.
For instance, a larger diameter represents a larger deceleration is desired.
\textcolor{black}{By presenting a graduated advisory scale—rather than a binary stop/go signal—the warning system reduces unnecessary hard alerts while still drawing attention to potential violations and let drivers to calibrate their own deceleration pattern. Meanwhile, to build user trust and adaptability, a feature that allows drivers to enable or disable the system can also be included to match individual preferences.}

\textcolor{black}{
We have chosen a colored structure for advising the human driver as empirical evidence strongly endorses the use of color-coded visual warnings in human-machine interaction (HMI). Use of color coding significantly improves drivers’ message comprehension, detection time, and reaction time compared to text-only signs according to a study by the U.S. Federal Highway Administration~\cite{kissner2021colorcms}. Moreover, a driving simulator experiment on 45 participants has shown that the red-themed message results in the fastest reaction time~\cite{friedrich2022urgency} leading us to use this color for the immediate and hard brake suggestion.}

Table~\ref{rlrws-warningmess} explains the interpretation of colors within the intensity range (0 to 100).
Green represents normal driving behavior, indicating that no specific guidance from the RLRWS is required, and encompasses all positive acceleration values along with minor deceleration values.
Yellow indicates the need for moderate deceleration.
Red denotes a potential risk of a red light violation and requires intense braking.
Typically, the red signal appears when the driver has not followed the earlier yellow warning.

\begin{table} 
    \centering
    \caption{Warning signal visualization}

    \begin{tabular}{ccc}
        \toprule
        \textbf{Color} & \textbf{Definition}  & \textbf{Range} \\
        \midrule
       Green     & {Normal driving behavior.} & below 10  \\ \hline
        Yellow        & {Decelerate gradually and a full stop is required soon.}  & between 10 and 60\\ \hline
        Red       &  {A hard brake is required to avoid running the red light.}& above 60 \\
        \bottomrule
        \label{rlrws-warningmess}
    \end{tabular}
    
\end{table}

\textcolor{black}{This variable-colored circle message is designed to enhance clarity for drivers:} (1) The adaptive sizing of the circles guides drivers regarding the suggested intensity for braking, allowing them to calibrate their own brake pedal response; (2) The use of color conveys the urgency of the message (e.g., red means an urgent braking situation); (3) The circular shape can potentially differentiate red light warning messages from other indicators in the vehicle (e.g. bars on the odometer), facilitating a quick response.
The visualization output updates upon receiving the latest warning signal.

\subsection{\textcolor{black}{Driver Acceptance and Human-Machine Interaction Considerations}}
\textcolor{black}{
A human driver can never be expected to follow the framework's warning closely. On the other hand, there is a direct relation between system effectiveness and driver acceptance as studied in~\cite{delaigue2008measuring}. Regarding driver acceptance, research studies have observed a positive attitude in drivers using advisory speed systems~\cite{etika2022predicting}. }

\textcolor{black}{
Our proposed framework is designed to be an advisory system in order to ensure a larger population of drivers will accept it. Comparing mandatory and advisory systems shows that although mandatory systems reduced speeds most, adaptive adjustments with advisory prompts was better accepted by drivers~\cite{delaigue2008measuring}. }

\textcolor{black}{
Analyzing video-recordings of 1,432 car trips shows that drivers naturally defer visual-manual tasks—such as dialing or reading reminders—when approaching risky contexts like turns or higher speeds, but may not allocate adequate safety margins~\cite{tivesten2015driving}. This supports our system’s design tenet: warnings should be context-triggered and minimal—provided only when risk is imminent—so as to nudge behavior without adding distraction. Finding the optimal time and method of visualizing the warning message is beyond the scope of this study and will be referred to as a future work consideration.}

\subsection{Numerical Solution}
In this work, the entire algorithm is implemented in Python.
The proposed traffic prediction algorithm predicts the traffic states for the road section \SI{500}{\meter} ahead of the ego vehicle over the next \SI{10}{\second}.
The length for each cell $dx$ is set to \SI{20}{\meter}.
The discretization time $dt$ in the traffic prediction is set to \SI{0.1}{\second}.
The algorithm updates the predicted traffic states every \SI{0.2}{\second} using the latest real-time traffic data.

\textcolor{black}{Based on previous studies on vehicle longitudinal motion optimization \cite{shao2021energy, he2025connectivity}, the constants use in the car-following constraints and optimization horizon are determined. Specifically, the optimization horizon $t_f$ for the proposed MPC-based optimization problem is set to \SI{10}{\second}.
To ensure reliable performance, the weighting factors in the MPC objective function are selected through an empirical tuning process, in which parameters are iteratively refined to achieve optimal control behavior.
}
To get its numerical solution, the Euler method with a $dt=$ \SI{0.2}{\second} discretization time step is used to balance computational burden and accuracy. 
The MPC problem is solved using IPOPT~\cite{biegler2009large} with CasADi~\cite{andersson2019casadi} as the modeling language.
The algorithm updates the optimal warning signal every \SI{1}{\second}.
On a MacBook Pro equipped with an M2 Max chip and 32 GB of RAM, each cycle of traffic prediction and optimization requires, on average, only \SI{0.45}{\second}, confirming the approach’s suitability for real‐time deployment.

\section{Simulation Results}\label{sec:sim-results}

\begin{figure}
    \centering
    \includegraphics[width=0.41\linewidth]{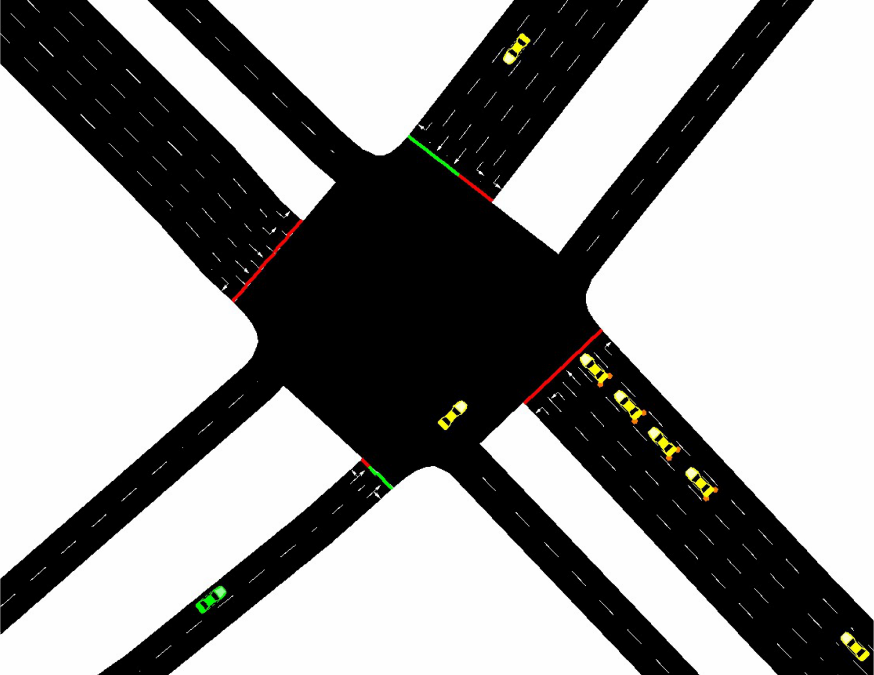}
    \caption{Modeled network with the ego vehicle shown in green and all other vehicles in yellow.}
    \label{fig:modeled-intersection}
\end{figure}

\textcolor{black}{We validate the proposed warning system in both simulation and on public road tests. Simulations admit a much greater range of scenarios and can test the system's guidance without being concerned for safety. During public road testing, the safety of the driver limits the range of driver behaviors, but the guidance can be tested with actual vehicles and traffic signals. This section presents our simulation results.}
The microscopic traffic simulator Simulation of Urban MObility (SUMO)~\cite{krajzewicz2002sumo} with the Krauss model as the car-following model~\cite{krauss1998microscopic} and a time step of \SI{0.1}{\second} is used for the numerical simulations. By default, SUMO will not allow a vehicle to violate a red light or even behave as a red light running vehicle.
\textcolor{black}{Thus, the SUMO driver model was modified so that the ego vehicle would run red lights (an abnormal behavior) unless stopped by the RLRWS.}
The four-direction multi-lane intersection (same intersection as the on-road testbed location in Section~\ref{sec:rlrws-roadtest-results}) at Scott County’s CSAH 18/CSAH 21/Southbridge Boulevard, Minnesota is modeled as the traffic network for the simulation, because it is also the intersection used for the public road testing.
Figure \ref{fig:modeled-intersection} depicts the geometry of the target intersection in the microsimulation environment.
The target intersection is a rural high-speed signalized intersection.
As observed in previous research, the dilemma zone issue at higher-speed signalized intersections poses greater challenges due to the wide range of vehicle speeds and the extended length of the type II dilemma zone, increasing the risk of severe collisions, particularly those involving heavy-duty trucks.
This makes the task of preventing red light violations not only more difficult but also critical for improving transportation safety.

In all numerical simulations, the stop bar of the intersection is located at \SI{0}{\meter}.
The ego vehicle is assumed to receive real-time SPaT information once it is within \SI{500}{\meter} of the intersection.
The Mersenne Twister algorithm~\cite{matsumoto1998mersenne} used by SUMO enables the randomness of the simulations.
Traffic Control Interface (TraCI) provided by SUMO is used to control the ego vehicle's movement.
Unless otherwise stated, it is assumed that the ego vehicle responds accurately to the optimal warning signal based on the driver model \eqref{eq:rlrws-mpc-driver-model} shown in Section\ref{sec:rlrws-driver-model} for our proposed warning system.
The ego vehicle's speed, acceleration, trajectory information, and signal status for both our proposed warning system and the baseline method are shown for each scenario. \textcolor{black}{Table \ref{tableparams} represents parameter values that are used in the simulation.}

\begin{table}[htp!]
\centering
\caption{\textcolor{black}{Traffic modeling, optimization, and simulation parameters simulated in results}}
\label{tableparams}
\renewcommand{\arraystretch}{1.3} 
\begin{tabular}{ | >{\centering\arraybackslash}m{2.5cm} | >{\centering\arraybackslash}m{2.5cm} | >{\centering\arraybackslash}m{2.5cm} | >{\centering\arraybackslash}m{2.5cm} | >{\centering\arraybackslash}m{2.5cm} |} 
  \hline 
  $v_0$ (m/s) & $a_\text{max}$ (m/s²) & $b$ (m/s²) &  $\tau$ (s) & c \\ 
  \hline
  24.6 & 2.6 & 4.5 & 1 & 10.14   \\ 
  \hline

  $h_\text{min}$ (s)  & $d_\text{min}$ (m) & $d_\text{max}$ (m) & $\text{gap}_\text{min}$ (m)  & $\rho_\text{jam}$ (v/km) \\ 
  \hline
  1.5 & 2.5 & $5 v_0$ & 2.5 & 130 \\ 
  \hline
\end{tabular}

\end{table}

In Section~\ref{sec:sim-no-preceding}, only the ego vehicle is simulated approaching a signalized intersection.
In Section~\ref{sec:sim-preceding-veh}, the ego vehicle is simulated to follow several vehicles toward a signalized intersection. 
For all scenarios in this subsection, it is assumed that the ego vehicle is equipped with onboard sensors (such as radar or camera, common in most latest manufactured vehicle) and can obtain its immediate preceding vehicle's speed and position information for use in the traffic prediction algorithm.
The remaining vehicles ahead of the immediate preceding one in Section~\ref{sec:sim-preceding-veh} are assumed to be non-connected vehicles, and their information remains unknown to the traffic prediction algorithm.
It should be noticed that if connected vehicles are present ahead of the immediate preceding vehicle within communication range, additional traffic information can be used to enhance the accuracy of traffic prediction results, improving the performance of the proposed warning algorithm.
Therefore, the results shown in this subsection represent the performance of our proposed RLRWS with the minimal amount of available information.

\subsection{Baseline RLRWS}\label{sec:rlrws-background-baseline}
We compare our results against a baseline RLRWS, 
the single-stage in-vehicle warning system from \cite{gelbal2020hardware}.
This approach, based on V2I communication, has been validated through both driving simulator tests and field experiments in previous research.

After the ego vehicle reaches the communication range of the RSU, the warning system uses obtained real-time SPaT data to calculate the remaining time $t_\text{rem}$ before the onset of the red signal.
The time for the ego vehicle to reach the intersection is then computed by the following equation:
\begin{equation}
    t_\text{veh}=d_\text{int}/v_\text{veh},
\end{equation}
where $d_\text{int}$ is the longitudinal distance between the ego vehicle and the upcoming signalized intersection, $v_\text{veh}$ indicates the ego vehicle's instantaneous speed.
If this time $t_\text{veh}$ is greater than the remaining green and yellow phases $t_\text{rem}$, a warning signal is presented to the driver.

\subsection{Simulations: Without Preceding Vehicle}\label{sec:sim-no-preceding}
We first compare the proposed RLRWS and baseline approach as the ego vehicle approaches the signalized intersection alone. 
Three scenarios are considered based on the traffic signal status and the ego vehicle's response:
First, the ego vehicle approaches the intersection with the traffic light already red.
Second, the ego vehicle approaches the intersection as the traffic light changes from green to red.
Third, the ego vehicle approaches the intersection when the traffic light is red, but the driver initially disregards the warning signal's guidance.
This specific scenario shows the robustness of the proposed algorithm in cases where the driver does not immediately respond to the warning signal.

\textcolor{black}{In all simulations, the trajectory, speed, acceleration of the ego vehicle, traffic signal status are shown, alongside the warning signals generated by both our proposed warning system and the baseline approach. In the first sub-figure, the green, yellow and red colors represent the instantaneous traffic signal status. In the first three sub-figures, black lines present vehicle dynamics controlled by the SUMO default controller, while blue lines show those controlled by the proposed algorithm. The last sub-figure displays the warning signals of our system and baseline approach.}

\subsubsection{Traffic Light Is Already Red}\label{sec:sim-1-results}

\begin{figure}
    \centering
    \includegraphics[width=0.6\linewidth]{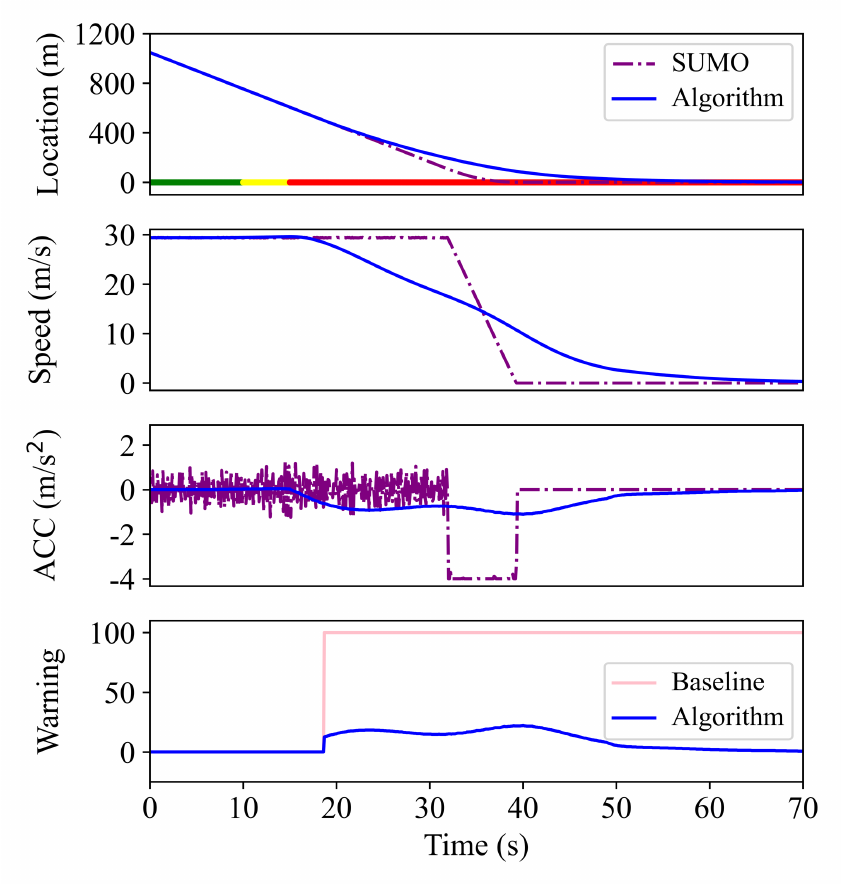}
    \caption[Comparison between the proposed warning algorithm and the baseline approach for simulation 1.]{\textcolor{black}{Comparison between the proposed warning algorithm and the baseline approach for simulation 1. In the first three sub-figures, blue and purple lines indicate vehicle's states using the proposed algorithm and SUMO default controller, respectively. The last sub-figure
displays the warning signals of our system and baseline approach.}}
    \label{fig:sim-single-1}
\end{figure}

When the ego vehicle approaches a signalized intersection during a red light, it is expected to stop completely before the stop bar.
In Figure \ref{fig:sim-single-1}, under the guidance of our proposed warning algorithm, the ego vehicle decelerates gradually and comes to a complete stop before the stop bar. 
\textcolor{black}
{In contrast, under identical conditions, the vehicle trajectory using the SUMO default controller exhibits noticeably less smooth maneuvers.
Compared to this, the proposed system reduces the peak deceleration during the slowing-down phase by 72.2\%, indicating significantly smoother driving behavior and a reduced risk of red-light violations.
This is attributed to the proposed system's ability to optimize vehicle trajectory using SPaT information from V2I communication.}
Such improvements not only enhance the comfort of the vehicle's operation but also hold the potential to increase energy efficiency and safety.

Meanwhile, compared to the step warning signal employed by the baseline approach, our method generates a more moderate warning signal.
This reduces the risk of aggressive braking often triggered by the abrupt onset of a single-stage warning signal, as noted in previous research.
Consequently, this enhancement further improves the ego vehicle's comfort and reduces the risk of rear-end collisions.

\subsubsection{Traffic Light Changes to Red}\label{sec:sim-2-results}

In the second scenario, the ego vehicle approaches the intersection as the traffic light changes from green to red, a scenario that commonly leads to red light violations, as drivers may be unaware of the impending red light.
This is particularly problematic at rural high-speed intersections where driver reaction times are slower and yellow intervals may not provide sufficient time to brake and stop safely, increasing the risk of collisions at the intersection. The intersection used in this study is a high-speed intersection in a suburb far from the city, on a road with relatively few traffic signals, and therefore a possible example of slow driver response.

\begin{figure}
    \centering
    \includegraphics[width=0.6\linewidth]{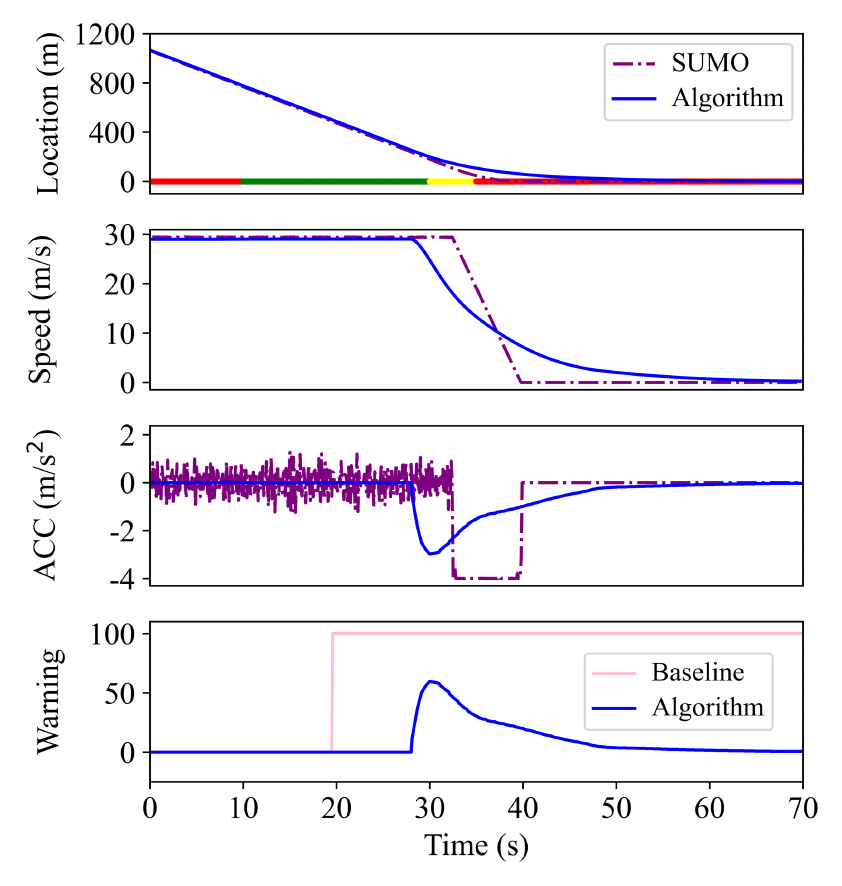}
    \caption[Comparison between the proposed warning algorithm and the baseline approach for simulation 2.]{\textcolor{black}{Comparison between the proposed warning algorithm and the baseline approach for simulation 2. In the first three sub-figures, blue and purple lines indicate vehicle's states using the proposed algorithm and SUMO default controller, respectively. The last sub-figure
displays the warning signals of our system and baseline approach.}}
    \label{fig:sim-single-2}
\end{figure}

As shown in Figure~\ref{fig:sim-single-2}, under the guidance of our proposed warning system, the ego vehicle begins decelerating even before the yellow light appears, allowing for a longer distance remained for braking.
Compared to the ego vehicle's trajectory under the same conditions using the SUMO default controller, the ego vehicle equipped with the proposed algorithm again exhibits smoother maneuver.
This improved performance is particularly beneficial for safety in adverse weather conditions, such as snowy days, when longer braking distances are required.

Based on Figure~\ref{fig:sim-single-2}, it is observed that the baseline approach triggers the single stage warning signal as soon as the vehicle enters the RSU's communication zone.
In contrast, our proposed system generates a timely and appropriate warning signal only when necessary.
This distinction is significant, as the abrupt onset of the step signal in the baseline method may lead to unnecessary deceleration, increasing the risk of rear-end collisions and disrupting traffic flow by creating shockwaves.
Additionally, since the warning signal's status remains unchanged, it offers no useful guidance for a smooth deceleration pattern.

\subsubsection{Driver Ignores RLRWS Guidance}\label{sec:sim-3-results}

As noted earlier, there is a significant possibility that drivers may ignore the warning signal, opting instead to maintain speed until the last moment before executing a sudden and abrupt deceleration.
Therefore, in the third scenario, we use the same driving condition as the first scenario to show the performance of our proposed algorithm under such driving patterns, where the driver ignores the warning signal, continues at a constant speed, and only decelerates when the ego vehicle nears the stop bar.
For this scenario, the red line in the third sub-figure shows the desired acceleration computed by the proposed algorithm, while the blue line represents the vehicle's actual acceleration.

\begin{figure}
    \centering
    \includegraphics[width=0.6\linewidth]{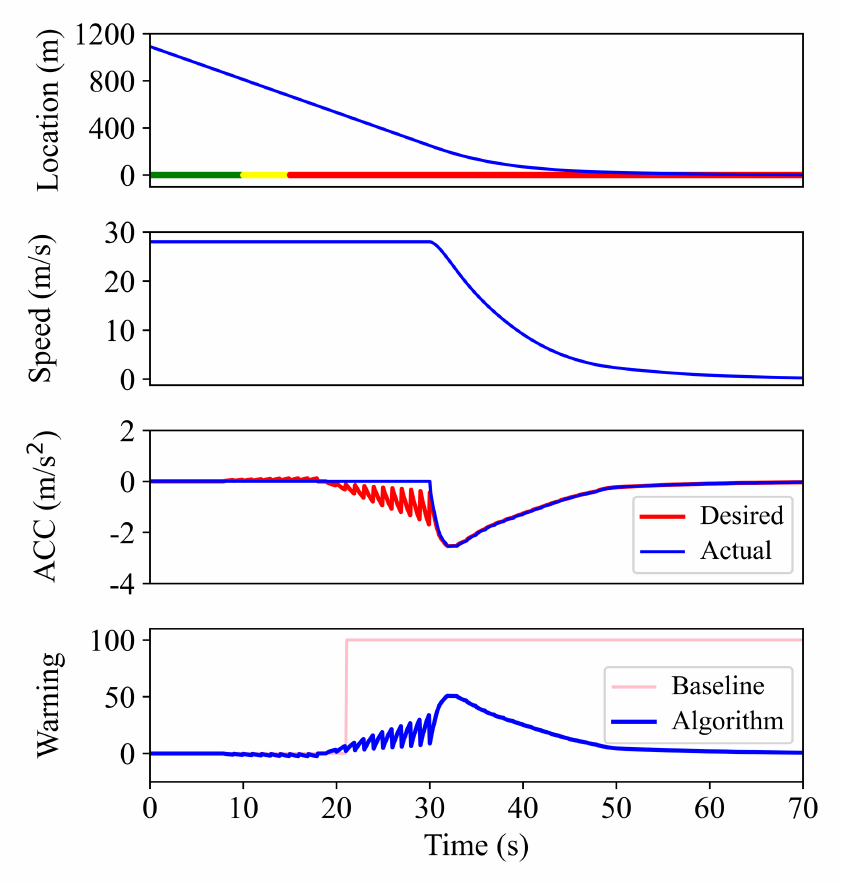}
    \caption[Comparison between the proposed warning algorithm and the baseline approach for simulation 3.]{\textcolor{black}{Comparison between the proposed warning algorithm and the baseline approach for simulation 3. In the first three sub-figures, blue lines indicate vehicle's states using the proposed algorithm.
    Red lines show the states that the driver initially fails to follow the warning signal's guidance. The last sub-figure
displays the warning signals of our system and baseline approach.}}
    \label{fig:sim-single-3}
\end{figure}


In Figure \ref{fig:sim-single-3}, the ego vehicle initially ignores the warning signal and fails to slow down.
As it nears the stop bar, our proposed system generates a larger warning signal, indicating the need for larger deceleration to prevent a red light violation.
While the baseline approach provides a correct warning signal once the vehicle enters the RSU's communication zone, its unchanged single stage warning signal may continue to be overlooked by the driver.
In contrast, our system adjusts the warning signal, shifting from yellow to red with an increasing circle diameter.
These changes in both color and size can effectively capture the driver's attention, offering more effective guidance to avoid red light violations.

\subsection{Simulations: With Preceding Vehicle}\label{sec:sim-preceding-veh}
We now study the behavior of the ego vehicle under the RLRWS as it approaches the signalized intersection behind a platoon of vehicles. 
Three driving scenarios are analyzed:
First, the platoon of vehicles approaches the intersection with the traffic light already red.
All vehicles must slow down and stop before the stop bar.
Second, the platoon of vehicles approaches the intersection as the traffic light changes from green to red.
The ego vehicle's immediate preceding vehicle can pass the intersection by the end of the yellow interval, but the ego vehicle must stop to avoid running the red light.
Third, the platoon of vehicles approaches the intersection as the traffic light changes from red to green.
However, there is a queue ahead of the intersection. 
All vehicles must slow down initially and then re-accelerate to pass the intersection.

\textcolor{black}{In the results, all vehicles' trajectories, the ego vehicle's and its immediate preceding vehicle's speed, acceleration, and traffic signal status are shown in the first three sub-figures.
The warning signals generated by both our proposed warning system and the baseline approach is presented in the last sub-figure.
In the first sub-figure, the green, yellow and red colors represent the instantaneous traffic signal status.
}

\subsubsection{Platoon Approaches a Red Traffic Light}

\begin{figure}
    \centering
    \includegraphics[width=0.6\linewidth]{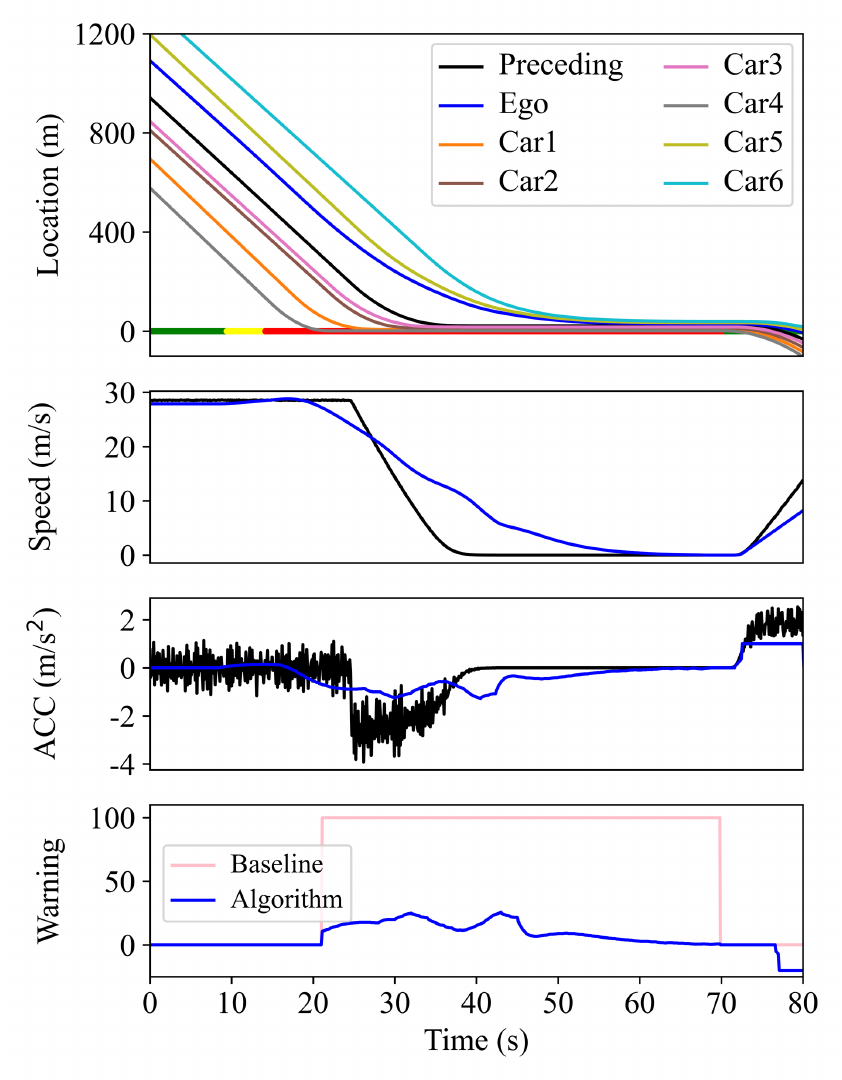}
    \caption[Comparison between the proposed warning algorithm and the baseline approach for platoon scenario 1.]{\textcolor{black}{Comparison between the proposed warning algorithm and the baseline approach for platoon scenario 1. In the first sub-figure, ground truth trajectories from different vehicles are shown in lines with different colors.
    In the first three sub-figures, blue lines indicate vehicle's states using the proposed algorithm.
    Black lines indicate that of the preceding vehicle.
    The last sub-figure
displays the warning signals of our system and baseline approach.}}
    \label{fig:sim-platoon-1}
\end{figure}

In this scenario, all vehicles are expected to stop completely before the stop bar.
Figure~\ref{fig:sim-platoon-1} shows the trajectories of the ego vehicle as well as the platoon of vehicles.
As illustrated in Figure~\ref{fig:sim-platoon-1}, under the guidance of our proposed warning system, the ego vehicle gradually slows down and stops before the stop bar.
In contrast to its immediate preceding vehicle, the ego vehicle exhibits smoother maneuver and uses less deceleration, thanks to the support of our warning system, resulting in increased spacing between these two vehicles.
This improves the safety of the ego vehicle.
Furthermore, when comparing the warning signals generated by the baseline approach and our method, it is evident that our system's the moderate and gradually changing warning signal can effectively mitigate the risk of abrupt braking, which can be triggered by the sudden appearance of a step warning signal.
This not only enhances the efficiency of the traffic flow but also improves overall safety.

\subsubsection{Platoon Approaches as the Traffic Light Changes from Green to Red}

In this scenario, the platoon of vehicles approaches the signalized intersection as the traffic light changes from green to red.
This situation poses a risk of red light violations, particularly if the ego vehicle's immediate preceding vehicle crosses the stop bar during the yellow interval, leaving insufficient time for the ego vehicle to also pass safely before the light turns red.
Consequently, the driver may not have enough time to react, potentially resulting in entering into the intersection after the light has turned red.


\begin{figure}
    \centering
    \includegraphics[width=0.6\linewidth]{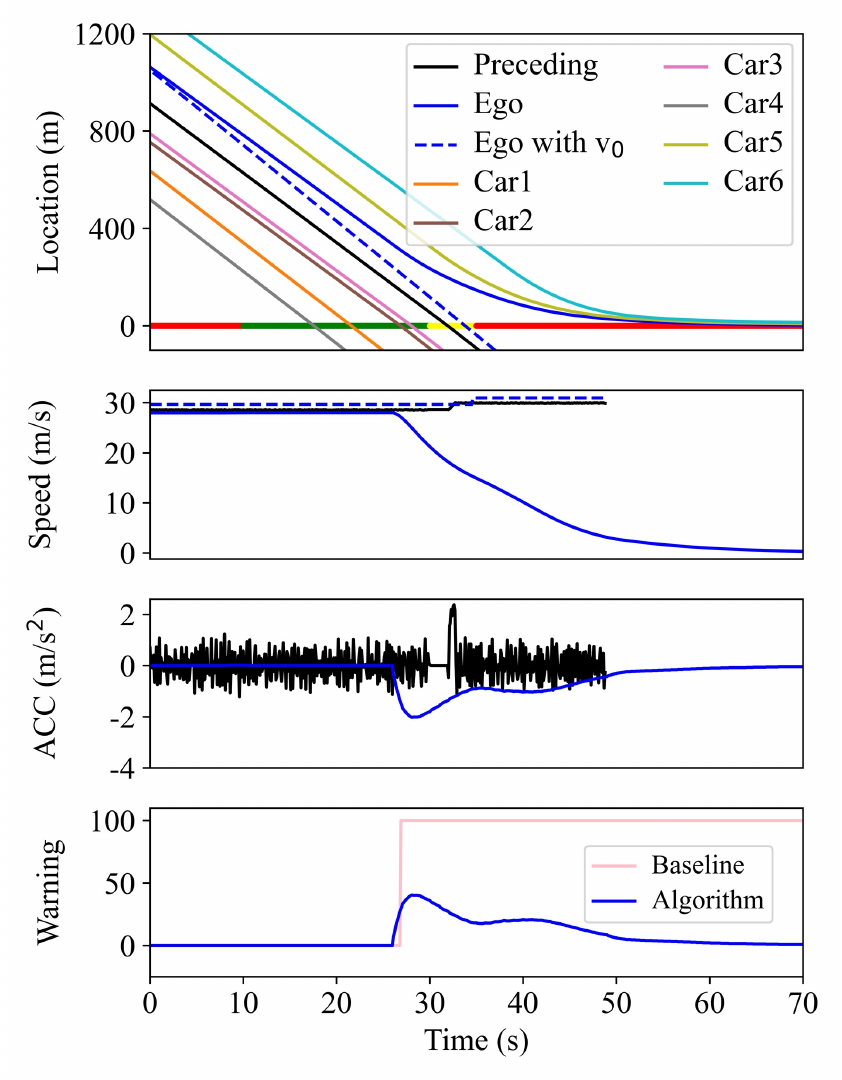}
    \caption[Comparison between the proposed warning algorithm and the baseline approach for platoon scenario 2.]{\textcolor{black}{Comparison between the proposed warning algorithm and the baseline approach for platoon scenario 2. In the first sub-figure, ground truth trajectories from different vehicles are shown in solid lines with different colors.
    The predicted trajectory of the ego vehicle using free flow speed $v_0$ is shown in blue dashed line.
    In the first three sub-figures, blue lines indicate vehicle's states using the proposed algorithm.
    Black lines indicate that of the preceding vehicle.
    The last sub-figure
displays the warning signals of our system and baseline approach.}}
    \label{fig:sim-platoon-2}
\end{figure}

In Figure~\ref{fig:sim-platoon-2}, if the preceding vehicles were not present, the ego vehicle would be able to pass the intersection before the end of the yellow interval.
However, in the simulated scenario where it follows a platoon of vehicles, the ego vehicle must decelerate to maintain a safe following distance from its immediate preceding vehicle.
Consequently, it is unable to cross the intersection prior to the light turning, influenced by the behaviour of these preceding vehicles.
In this context, guided by our proposed warning system, the ego vehicle gradually reduces its speed in advance, thereby successfully avoiding a red light violation.

Furthermore, upon comparing the warning signals generated by the baseline approach with those produced by our method, it is evident that the baseline warning system fails to deliver an appropriate warning in this scenario.
The step warning signal only activates after the ego vehicle has already begun to slows down under the guidance of our warning system.
However, without our proposed warning system, the vehicle may not decelerate at all, leaving the baseline approach ineffective in providing any warning due to its lack of traffic prediction capabilities ahead of the ego vehicle.
Consequently, this could lead to abrupt braking at the last moment or crossing the intersection after the traffic signal has turned red, both of which are dangerous situations.

\subsubsection{Ego Vehicle Approaches an Intersection with a Queue}

In the previous scenarios, there are no vehicle queues at the intersection.
However, in reality, due to the presence of a slow-moving queue, vehicles may need to slow down or stop even when the traffic light is green.
Depending on the queue's length, it takes some time for the intersection to clear, and new arriving vehicles must join the slower-moving queue until it disperses.
In this scenario, the platoon of vehicles approaches the intersection as the traffic light changes from red to green. 
However, because of the queue, they must initially slow down before safely passing through the intersection.

\begin{figure}
    \centering
    \includegraphics[width=0.6\linewidth]{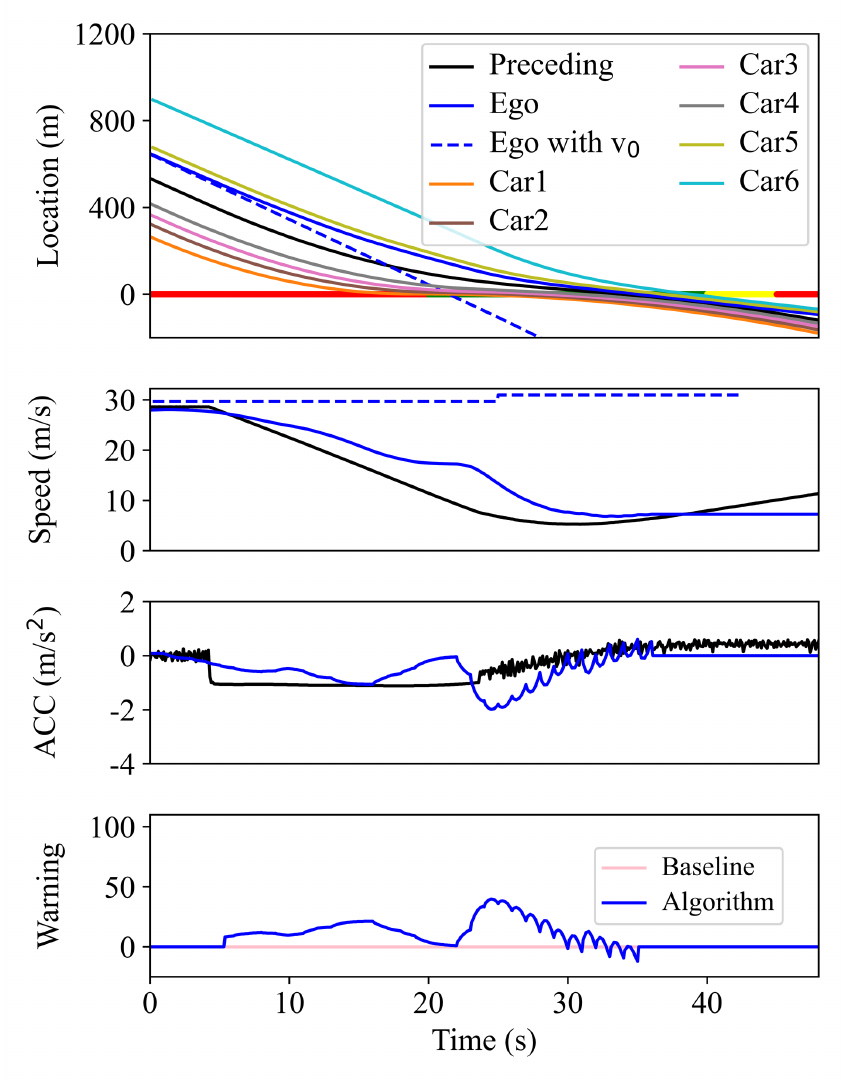}
    \caption[Comparison between the proposed warning algorithm and the baseline approach for platoon scenario 3.]{\textcolor{black}{Comparison between the proposed warning algorithm and the baseline approach for platoon scenario 3. In the first sub-figure, ground truth trajectories from different vehicles are shown in solid lines with different colors.
    The predicted trajectory of the ego vehicle using free flow speed $v_0$ is shown in blue dashed line.
    In the first three sub-figures, blue lines indicate vehicle's states using the proposed algorithm.
    Black lines indicate that of the preceding vehicle.
    The last sub-figure
displays the warning signals of our system and baseline approach.}}
    \label{fig:sim-platoon-3}
\end{figure}


Figure~\ref{fig:sim-platoon-3} shows that all vehicles slow down even as the traffic light changes from red to green.
In this specific scenario, although the prediction algorithm of our proposed warning system lacks information about the queue's existence, the warning system still guides the ego vehicle to gradually slow down gradually using the immediate preceding vehicle's information.
This results in a smoother maneuver for the ego vehicle.
In contrast, the baseline approach can not provide any meaningful guidance, as it fails to predict traffic conditions and consistently assumes that the ego vehicle can pass the intersection at a constant speed during the green interval.

Notice that other scenarios, such as when traffic in front of the ego vehicle is not traveling at free flow speed, including a slower truck passing through the intersection or a vehicle stopping at a green light to yield for a left turn, can be similarly represented and modeled.
In these scenarios, the ego vehicle will also slows down gradually in advance and pass the intersection safely at a reduced speed under the guidance of the proposed algorithm.

\section{Experiments on Public Roads}\label{sec:rlrws-roadtest-results}

To demonstrate the performance of the proposed warning algorithm in real-world traffic scenarios, road tests are conducted using the testbed configured by the research group.
Three different examples are analyzed using ego vehicle's trajectory, speed and warning signal information, including data generated by both our proposed warning system and the baseline approach.
When the RSU is within the OBU's communication range, the traffic signal status is also included in the results plotting.
For the trajectory visualization, the stop bar is located at \SI{0}{\meter}.

During the road tests, the ego vehicle operates within typical traffic conditions, surrounded by numerous vehicles exhibiting diverse and unknown driver behaviours.
However, only the ego vehicle's information and real-time SPaT information from the RSU are available to our proposed warning system.
This presents a particularly challenging scenario, as the information of the immediate preceding vehicle is also unavailable, limiting the system's performance.
Nonetheless, the results indicate that our system consistently provided reasonable outputs even in these challenging tests.

\begin{figure}
    \centering   
    \includegraphics[width=0.7\linewidth]{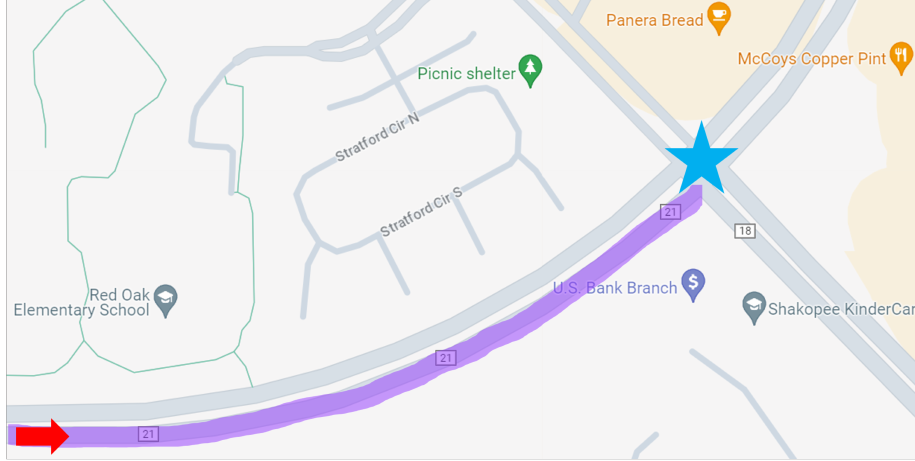}
    \caption{Map of the intersection equipped with the RSU.}
    \label{fig:rlrws-map-intersection}
\end{figure}

\subsection{Configuration of On-Road Testbed} \label{sec:rlrws-on-road-testbed}
\begin{figure*}
\begin{subfigure}[t]{0.45\textwidth}
{\includegraphics[width=0.65\linewidth]{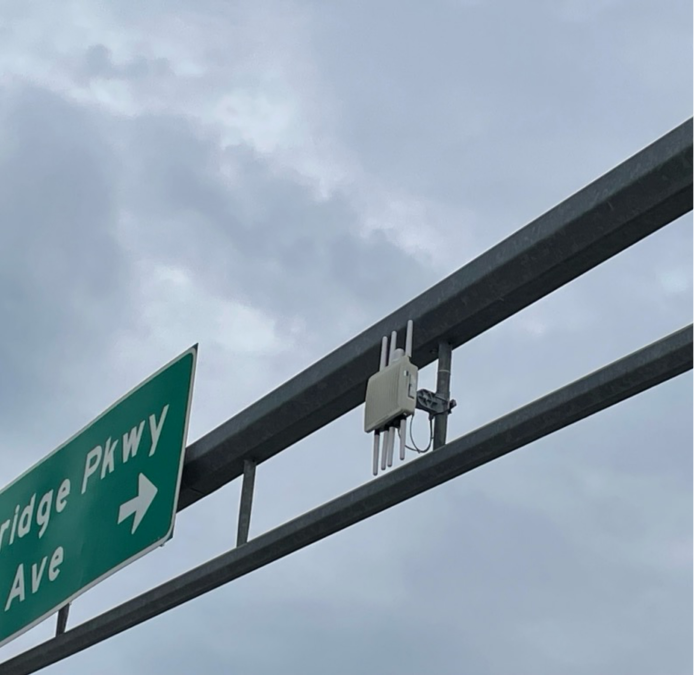}}
\caption{RSU installed at the intersection.}
\label{fig:rlrws-rsu}
\end{subfigure}
\begin{subfigure}[t]{0.45\textwidth}
{\includegraphics[width=0.65\linewidth]{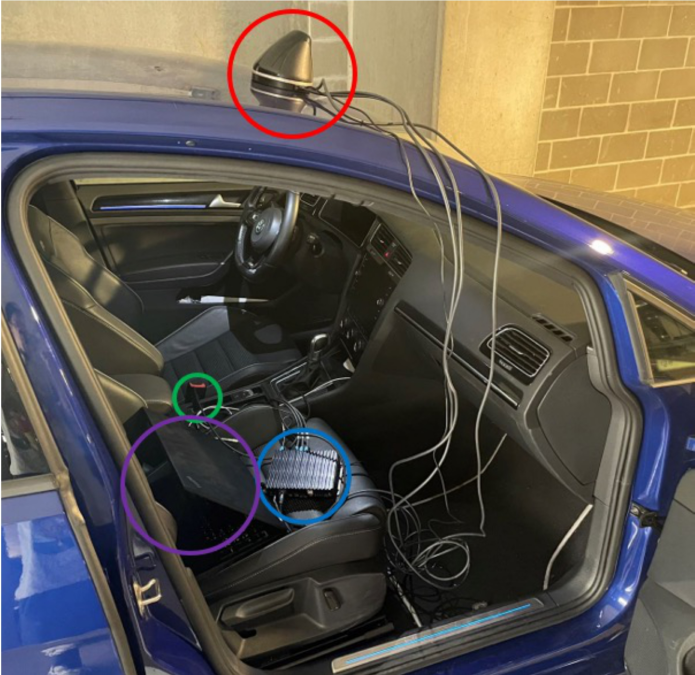}}
\caption{Test vehicle setup.}
\label{fig:rlrws-obu}
\end{subfigure}
\centering
\caption{RSU and OBU used for road tests.}
\label{fig:rlrws-hardware}
\end{figure*}

\begin{figure}
    \centering
    \includegraphics[width=0.5\linewidth]{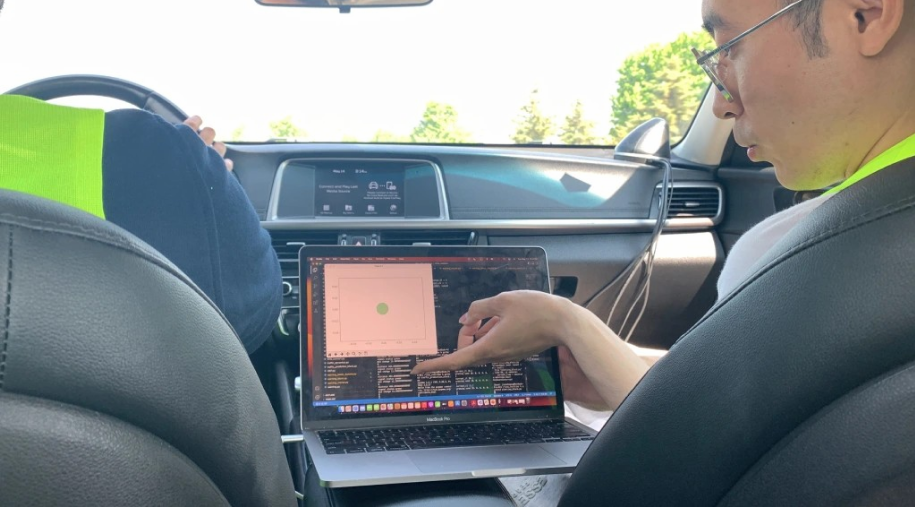}
    \caption{The in-vehicle warning system displays the warning signal during a road test.}
    \label{fig:rlrws-demonstration}
\end{figure}

\textcolor{black}{
An on-road testbed including test vehicles, OBU and RSU is configured for validating the proposed warning system~\cite{levin2024development}.
The RSU (Figure~\ref{fig:rlrws-rsu}), which broadcasts real-time SPaT information, is installed at the intersection of Scott County's CSAH 18/CSAH 21/Southbridge Boulevard in Minnesota (Figure~\ref{fig:rlrws-map-intersection}).
The purple sector in Figure~\ref{fig:rlrws-map-intersection} shows the road section used for later road tests and red arrow indicates the driving direction.
The position of the target intersection is marked with the blue star.}
\textcolor{black}{
The test vehicle (Figure~\ref{fig:rlrws-obu}) is equipped with an OBU connected to a laptop.
The OBU receive the teal-time SPaT data, along with the ego vehicle's speed and location information via its antenna, transmitting these to the laptop in real-time.
}
The laptop then processes the received information, calculates the optimal warning signal using our proposed algorithm, and displays the warning signal on its screen in real-time. 
Figure~\ref{fig:rlrws-demonstration} shows the warning signal displayed by the in-vehicle warning system during a road test. 
Meanwhile, links for videos of demonstration and more information are provided in the Acknowledgment section.

\textcolor{black}{In the road test, the ego vehicle operated by the research group is the only vehicle equipped with connectivity technology.
All surrounding vehicles are conventional human-driven vehicles without connectivity, so the ego vehicle has no access to their states and must cope with a wide range of unobservable driver behaviors.}
To ensure traffic safety during each experiment, the ego‑vehicle driver rigorously follows all applicable traffic laws and adapts to the surrounding traffic environment.
According to the previous research, the communication latency for the connected vehicle is around \SI{0.15}{\second}~\cite{orosz2016connected}.
Vehicle position and speed are obtained from the on‑board unit’s Global Navigation Satellite System (GNSS) based localization module, whose accuracy is better than \SI{0.1}{\meter}.


\subsection{Trajectory Results}

We now demonstrate our proposed warning system in real-time using the configured on-road testbed. 
During the road tests, the ego vehicle's dynamics, the optimized warning signal, and the traffic signal status are recorded. 
Meanwhile, to provide a comparison with the baseline approach, the warning signals generated by the baseline warning system are also included.

\textcolor{black}{
In the following subsections, the trajectory, speed profiles of the test vehicle, traffic signal status are shown, alongside the warning signals generated by both our proposed warning system and the baseline approach are shown for each scenario. In the first sub-figure, the green, yellow and red colors represent the instantaneous traffic signal status when the RSU is within the communication range of the test vehicle.
In the last sub-figure, the warning signals of our proposed system and baseline system are plotted in blue and pink, respectively.}

Three driving scenarios are studied to show the application of our proposed warning system in real-world traffic conditions:
In the first scenario, the ego vehicle approaches the intersection with the traffic light already red, and an unknown number of vehicles waiting before the stop bar.
In the second scenario, the ego vehicle approaches the intersection when the traffic light changes from green to red.
However, the driver initially fails to follow the guidance of the warning system and only begins decelerating after red warning signal is triggered, a typical scenario for red light violations.
In the third scenario, the ego vehicle approaches the intersection as the traffic light just turns red.
Once again, the driver ignores the initial warning signal, leading the system to issue a larger warning signal.
These scenarios demonstrate how our system adapts to real-world challenges, including driver behavior and traffic signal changes.

\subsubsection{Traffic Light is Already Red}

\begin{figure}
    \centering
    \includegraphics[width=0.6\linewidth]{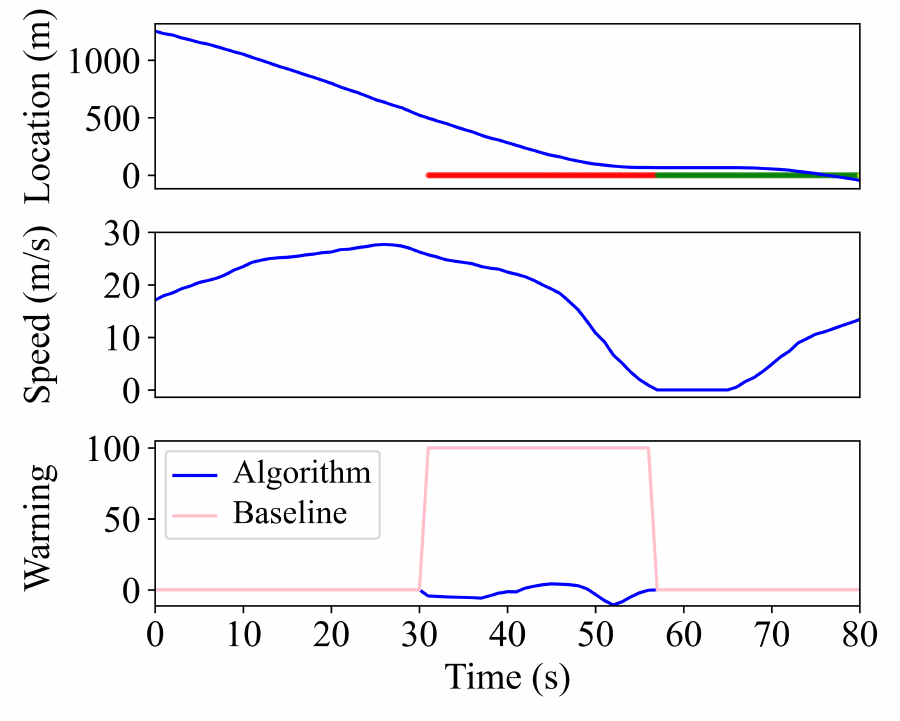}
    \caption[Comparison between the proposed warning algorithm and the baseline approach for scenario 1.]{\textcolor{black}{Comparison between the proposed warning algorithm and the baseline approach for scenario 1. Blue lines show the vehicle states using the proposed algorithm. The pink line in the last sub-figure shows the warning signal from the baseline approach.}}
    \label{fig:road-test-1}
\end{figure}

When the test vehicle approaches the target intersection during a red light, it is expected to stop completely before the stop bar.
%
In Figure~\ref{fig:road-test-1}, upon entering the RSU's communication zone, our proposed warning system begins to compute warning signal.
Since the traffic prediction algorithm does not have access to the preceding vehicle's information, it predicts the test vehicle can drive at the road section's free flow speed.
Therefore, the system generates a minor negative warning signal between \SI{30}{\second} and \SI{40}{\second} to guide the driver to speed up to catch the free flow speed and guarantee the traffic flow's efficiency.
In this period, since the warning signal is negative, a green message is displayed on the screen and will not influence the driver's normal driving behaviour.

As the test vehicle nears the intersection, the positive warning signal of our proposed system suggests a deceleration to the driver.
Since there is enough remaining distance for braking, the warning signal displayed on screen is only in yellow stage.
During this process, to maintain a safe spacing between the test vehicle and its preceding vehicle, the driver applies more brake force than the system suggests.
Consequently, around \SI{50}{\second}, the system generates another negative warning signal---a green message on the screen---advising the driver to resume normal driving.
This is because the system does not know the existence of the preceding vehicle and advises the driver to stop close to the stop bar.
This performance can be improved when the preceding vehicle's information is available through onboard sensors.

Meanwhile, compared to the step warning signal employed by the baseline approach, the more moderate warning signal generated by our proposed warning system can lead to smooth vehicle operation, potentially improving test vehicle's safety, energy efficiency and comfort.
This is consistent to the performance shown in the numerical simulations.

\subsubsection{Traffic Light Changes from Green to Red}

\begin{figure}
    \centering
    \includegraphics[width=0.5\linewidth]{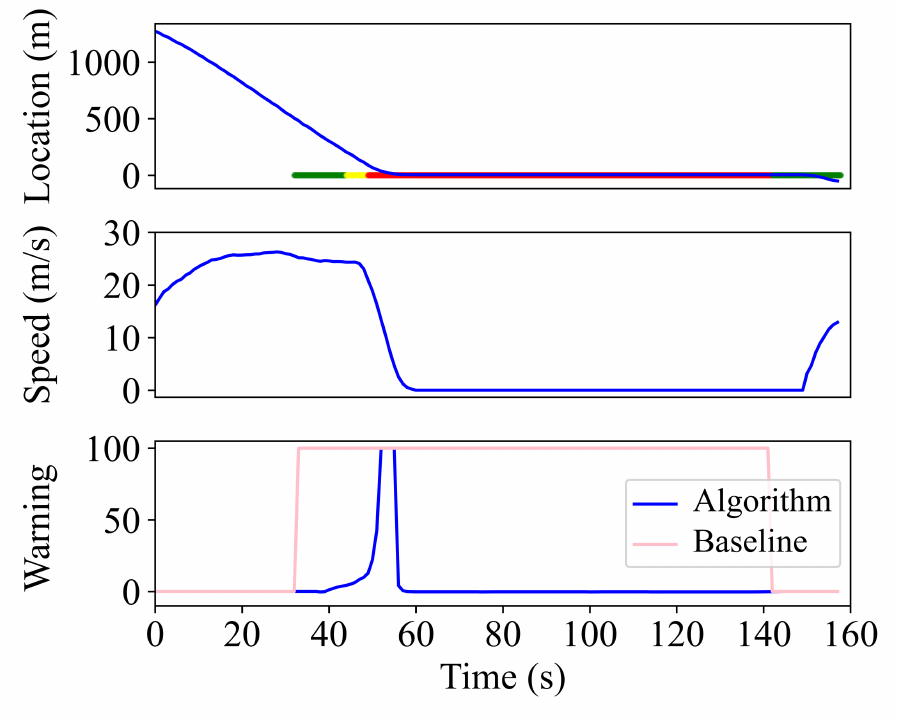}
    \caption[Comparison between the proposed warning algorithm and the baseline approach for scenario 2.]{\textcolor{black}{Comparison between the proposed warning algorithm and the baseline approach for scenario 2. Blue lines show the vehicle states using the proposed algorithm. The pink line in the last sub-figure shows the warning signal from the baseline approach.}}
    \label{fig:road-test-3}
\end{figure}

When the test vehicle nears the target intersection as the signal changes from green to red, it must slow down timely to avoid red light violation.
In Figure~\ref{fig:road-test-3}, upon entering the RSU's communication zone, our proposed warning system begins to alert the driver to slow sown.
However, the driver initially fails to brake mildly under the guidance of warning system and maintains the driving speed.
This results in a larger warning signal  around \SI{50}{\second} of the road test as the test vehicle gets closer to the stop bar.
During this period, the warning signal shown on screen changes from yellow to red with an increasing circle size.
This alerts the driver to slow down immediately to avoid running the red light.

When we compare the warning signal from our proposed system with that of the baseline approach, the single-stage warning signal can not attract the driver's attention timely as it appears for a long while.
Our proposed warning system's signal changes in both color and size, which can alert the driver more effectively when the driver initially ignores the warning system. Our warning signal could be augmented with audio cues too.

\subsubsection{Driver Ignores Initial RLRWS Guidance}

\begin{figure}
    \centering
    \includegraphics[width=0.5\linewidth]{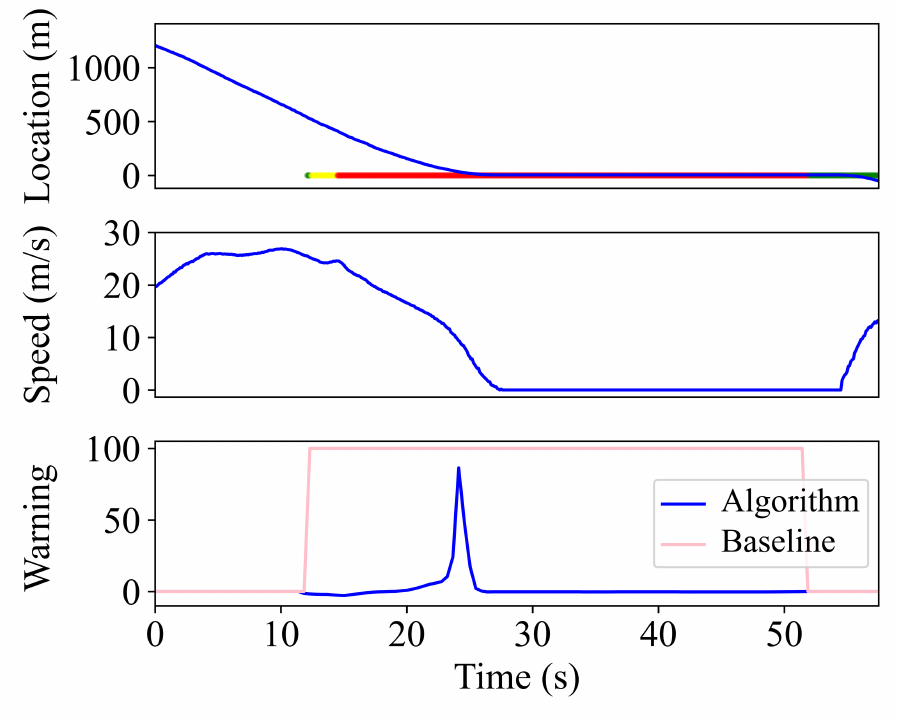}
    \caption[Comparison between the proposed warning algorithm and the baseline approach for scenario 3.]{\textcolor{black}{Comparison between the proposed warning algorithm and the baseline approach for scenario 3. Blue lines show the vehicle states using the proposed algorithm. The pink line in the last sub-figure shows the warning signal from the baseline approach.}}
    \label{fig:road-test-2}
\end{figure}

In the third example, the test vehicle approaches the intersection as the traffic signal just turns red, but the driver initially does not start braking or respond to the warning signal.
As shown in Figure~\ref{fig:road-test-2}, the warning system is triggered once the test vehicle enters the RSU' communication zone.
Since the driver initially does not slow down sufficiently as the proposed warning system's suggestion between the \SI{10}{\second} and \SI{20}{\second} of the road test, our warning system increases the warning signal's value and presents a bigger red circle to alert the driver decelerate immediately to stop before the stop bar.
Similar to the last example, compared to the step warning signal of baseline method, our proposed system still functions effectively under this kind of conditions.

\section{Conclusions and Future Work}\label{sec:conclusion}

\textcolor{black}{In this work, we introduced a novel traffic prediction based warning framework that provides in-vehicle tailored warning signals to individual drivers instead of issuing broadcast warnings.
The proposed system is based on vehicle connectivity, where real-time SPaT data and information from other connected vehicles, when they are available, are used by a traffic prediction algorithm to predict future traffic conditions towards the signalized intersection.
\textcolor{black}{The prediction results are used by a model predictive control based optimization problem that computes the optimal warning signal to guide the driver to slow down properly and avoid red-light violations.}
Then, a continuously updated in-vehicle display conveys that advice in a intuitive and graded format.
Since the system continuously updates the traffic prediction and warning signal as the ego vehicle approaches the intersection, changes in traffic conditions and driver's behavior to the previous warning signal are considered upon computing the new warning signal.}

\textcolor{black}{We validated the system across a broad spectrum of driving conditions—combining high‑fidelity simulations with real‑world road tests—and benchmarked it against a conventional in‑vehicle warning baseline. The proposed framework consistently outperformed the comparator, sidestepping several key shortcomings of earlier solutions. In simulation, it reduces the ego vehicle’s peak deceleration by up to 72.2\%, demonstrating its capacity to deliver timely guidance that enables drivers to brake smoothly and avoid red‑light violations.}

\textcolor{black}{Future work will employ data-driven methods to develop a more accurate driver behavior model. Historical data collected from road tests and driving simulators will be used to construct a nonlinear mapping between the displayed warning signal, traffic conditions, and the driver's response.}
\textcolor{black}{Meanwhile, it will comprise an extensive field campaign involving multiple connected vehicles in congested and uncongested roads, and considering varying signal timings.
Road tests will be conducted with a diverse range of drivers, including various age groups, levels of experience, and diverse driving behaviors.
Meanwhile, additional signalised intersections will be installed with RSUs, enabling the proposed algorithm to be validated under a wide range of traffic densities and environmental conditions.
These tests will evaluate the algorithm's long-term robustness and effectiveness, including comprehensive false-positive and false negative analyses of the proposed warning algorithm and any driver distraction the system may cause. Feedback will be used to refine the algorithm and improve the system’s accuracy and responsiveness.}

\section*{Acknowledgement}
The authors gratefully appreciate the support of the Minnesota Department of Transportation (report published (\cite{levin2024development}). All views expressed in this paper belong to the authors alone.

\bibliography{mybibfile}

\appendix

\section{Traffic prediction}
\label{sec:rlrws-background}

The decision of whether to stop or enter the intersection on yellow depends on the future trajectory of the ego vehicle, which in turn depends on the traffic conditions ahead of the ego vehicle. Prior work on in-vehicle warning systems assumed constant speed or deceleration by the ego vehicle, but that assumption is frequently wrong in the presence of other traffic~\cite{yan2015effect,johnson2019connected}.
\textcolor{black}
{We use traffic prediction, assuming that a subset of vehicles are equipped with V2V communications. 
The positions and number of the remaining human-driven vehicles, however, are not known to the algorithm.}
In this section, we briefly introduce the traffic prediction framework used in the proposed warning system.
The framework developed in \cite{shao2020eco} is used to predict the traffic conditions along the ego vehicle's driving lane within the algorithm's prediction horizon. \textcolor{black}{
As the key contribution of this work is the structure of the warning algorithm rather than traffic prediction, the impact of lane changing maneuvers on traffic prediction is not the focus, but lane changing can be incorporated into the traffic prediction. For interested readers, the algorithm shown in \cite{he2023real, zamanpour2025incorporating, he2025connectivity} can be easily adapted to the proposed warning system framework to incorporate the impact of lane changing maneuvers into the traffic prediction process.}

\begin{figure}
    \centering   
    \includegraphics[width=0.88\linewidth]{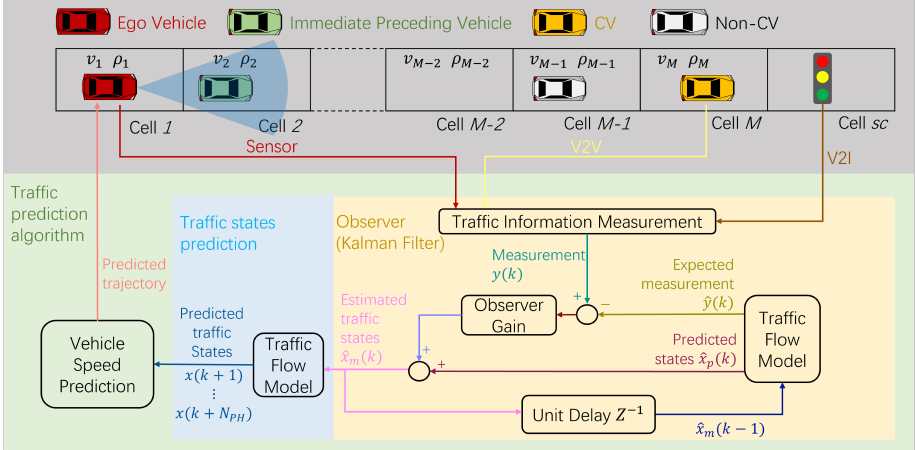}
    \caption{Traffic prediction framework used in this work.} 
    \label{fig:rlrws-prediction-structure}
\end{figure}

Figure~\ref{fig:rlrws-prediction-structure} shows the traffic prediction framework.
\textcolor{black}{
In the figure, the red vehicle signifies the ego vehicle equipped with the proposed algorithm.
Yellow vehicles indicate connected vehicles, while gray vehicles represent non-connected ones.}
\textcolor{black}{Equipped with an on‑board unit (OBU) and perception sensors, the ego vehicle receives real‑time inputs from V2V and V2I links, supplemented by its own camera–radar sensors. 
Because full state information for every vehicle upstream is unattainable—especially in mixed‑traffic platoons—microscopic car‑following models such as the IDM are unsuitable to predict the ego vehicle's future trajectory.
Therefore, we employ an unscented Kalman filter (UKF)~\cite{wan2000unscented} and a macroscopic traffic flow model to estimate unknown traffic conditions using partial measurements.
Then we can propagate the traffic flow model along the prediction horizon based on the estimated traffic states to predict future traffic conditions ahead of the ego vehicle.}

\textcolor{black}{The UKF can be used to estimate a system's states by using the system dynamic model together with the partial observation of the system. 
In the traffic prediction framework used in this work, the macroscopic traffic flow model is used as the system's dynamic model. 
The partial observations of the system include the real-time SPaT information from V2I communication, along with the speed and position data of the ego vehicle (via GPS antenna), immediate preceding vehicle (via onboard sensors), and connected vehicles (via V2V communication when they are in the communication range).
The UKF estimates the traffic conditions ahead of the ego vehicle in terms of traffic speed and traffic density (yellow part in the figure) by using the traffic flow model and partial observation of the traffic states in term of connected vehicle speed and location.
Consequently, the algorithm computes the future traffic conditions along the prediction horizon by propagating the traffic flow model (blue part in the figure).
Since the individual vehicle's driving speed correlates with the traffic speed at its location, the algorithm ultimately computes the ego vehicle and its immediate preceding vehicle's predicted longitudinal speed and trajectory.}

Specifically, the algorithm utilizes the discretized second-order Payne Whitham (PW) model~\cite{wang2005real} as the system's dynamic model, which is given by
\begin{subequations}
\setlength{\abovedisplayskip}{3pt}
\setlength{\belowdisplayskip}{3pt}
\begin{align}
    \rho_{j}(k+1) =& \rho_{j}(k) - \frac{dt}{d x} \left[ \rho_{j}(k)v_{j}(k) - \rho_{j-1}(k)v_{j-1}(k) \right] + \omega_{j}(k), \label{eq:rho}\\
     v_{j}(k+1) =& v_{j}(k) - \frac{dt}{d{x}}v_{j}(k)[v_{j}(k)-v_{j-1}(k)] 
      + \underbrace{dt \cdot \frac{[{V}_{e}(\rho_{j}(k))-v_{j}(k)]}{\tau}}_{\text{Speed adaptation}}-\\&\underbrace{\frac{dt}{d{x}} \cdot \frac{c_{0}^{2} \cdot [\rho_{j+1}(k)-\rho_{j}(k)]}{\rho_{j}(k)+\epsilon} }_{\text{Traffic pressure}}+ \xi_{j}(k),\label{eq:v}
\end{align}\label{eq:pw}
\end{subequations} \hfill \break
where \eqref{eq:rho} and~\eqref{eq:v} describe the evolution of traffic density and traffic speed, respectively.
$k$ is the discretized time instance;
$dt$ denotes the time step size;
$dx$ represents the length of each cell;
$j$ is the cell index;
$\epsilon$ is a small positive number to prevent zero denominator;
$\omega_{j}(k)$ and $\xi_{j}(k)$ describe model uncertainties assumed to follow a Gaussian distribution;
$c_{0}$ characterizes traffic pressure and $\tau$ describes the adaptation rate to reach the equilibrium speed;
${V}_e(\rho_{j}(k))$ signifies the equilibrium speed of cell $j$ and a triangular fundamental diagram is used to compute the equilibrium speed-density function in this work:
\begin{equation}
	V_{e}(\rho_{j}(k)) =
	\begin{cases}
		v_{0}, & 0 \leq \rho_{j}(k) \leq \rho_{c}    \\
		c \left(\dfrac{\rho_{\text{jam}}}{\rho_{j}(k)} - 1\right), & \rho_{c} < \rho_{j}(k) \leq \rho_{\text{jam}}
	\end{cases}   \label{eq:orignial-equilibrium}
\end{equation}
with
\begin{equation}
    \rho_{c} = \frac{\rho_{\text{jam}}}{v_{0}/c + 1}    \label{eq:rho-c}
\end{equation}
where $v_{0}$ and $c$ are the free flow speed and the slope of density drop when traffic is congested, respectively.
$\rho_{\text{jam}}$ is the jam density.
$\rho_{c}$ is the critical density given by \eqref{eq:rho-c}.

Supposing an intersection with the signal controller is located at cell $sc$, the speed of this cell is set to zero when the traffic light is red:
\begin{equation}
v_{sc}(k+1) = 
    \begin{cases}
	    0, & \text{signal is red} \\
		\text{right side of~\eqref{eq:v},} & \text{signal is green or yellow}
	\end{cases}\label{eq:red-spd}
\end{equation}

To link the traffic speed and individual vehicle's driving speed, the following approximation is used to calculate the $i$-th vehicle's speed as a linearly interpolated speed of two cells adjacent to this vehicle on its driving lane:
\begin{equation}
    y_i(k) = \alpha_i (k)v_{j_\text{adj}+1}(k)+(1-\alpha_i(k))v_{j_\text{adj}}(k) + \phi_i(k) \label{eq:spd-measure}
\end{equation}
where $j_\text{adj}$ is the index of the last cell that the vehicle has passed.
$\alpha_i(k) = d_i/dx - j_\text{adj}$ is the interpolation coefficient with $d_i$ as the location of the $i$-th vehicle.
$\phi_{i}(k)$ denotes measurement uncertainty following a Gaussian distribution.
This approximation is also employed to compute the ego vehicle's predicted speed along the prediction horizon.

\end{document}